# Non–Linear Clustering
# in the Cold+Hot Dark Matter Model

**Silvio A. Bonometto**[1,2], **Stefano Borgani**[3,4],
**Sebastiano Ghigna**[1,2], **Anatoly Klypin**[5] & **Joel R. Primack**[6]

[1]*Dipartimento di Fisica dell'Università di Milano*
*Via Celoria 16, I-20133 Milano, Italy*
[2]*INFN, Sezione di Milano,*

[3]*INFN, Sezione di Perugia,*
*c/o Dipartimento di Fisica dell'Università,*
*via A. Pascoli, I-06100 Perugia, Italy*

[4]*SISSA – International School for Advanced Studies,*
*via Beirut 2–4, I-34014 Trieste, Italy*

[5]*Astronomy Department, New Mexico State University*
*Box 30001 / Dept. 4500, Las Cruces, NM 88003-0001, USA*
*and Astro-Space Center, Lebedev Physical Institute, Moscow, Russia*

[6]*Institute for Particle Physics, University of California,*
*Santa Cruz, CA 95064, USA*

# Abstract

We use high resolution ($512^3$ grid points) particle–mesh (PM) N–body simulations to follow the development of non–linear clustering in a $\Omega = 1$ Universe, dominated by a mixture of Cold + Hot Dark Matter (CHDM) with $\Omega_{cold} = 0.6$, $\Omega_{hot} = 0.3$ and $\Omega_{baryon} = 0.1$; a simulation box of 100 Mpc a side ($h = 0.5$) is used. We analyze two CHDM simulations with present time corresponding to the linear biasing factor $b = 1.5$ (COBE normalization), starting from different initial random numbers. We also compare them with CDM simulations with $b = 1.5$ and $b = 1$ (COBE–normalized CDM).

We evaluate high–order correlation functions and the void–probability–function (VPF), compare them with observational data, and test models of non–linear clustering. Correlation functions are obtained both from counts in cells and counts of neighbors. Results are in close agreement and complement one another over different scale ranges. The analysis is made for DM particles and for galaxies, identified as massive halos in the evolved density field. We also check the effects of dynamical evolution and redshift space distortions.

We find that clustering of dark matter (DM) particles systematically exhibits deviations from the hierarchical scaling, $S_q \equiv \bar{\xi}_q/\bar{\xi}_2^{q-1} = $ constant, although the deviation decreases somewhat in redshift space. Galaxies follow hierarchical scaling far more closely, with coefficients $S_3 \sim 2.5$ and $S_4 \sim 7.5$, in general agreement with observational results. Unlike DM, the scaling of galaxy clustering is just marginally affected by redshift distortions and also does not sensitively depend on our choices for the initial spectra. The hierarchical scaling of galaxy clustering is confirmed by the VPF analysis. Again, in all the cases considered a good agreement with observational results is obtained.

The simultaneous use of cell and neighbor counts allows us to observe slight deviations of galaxy clustering from a scaling regime, however far less prominent than for DM. Such deviations are related to the amount of bias and also show features in analogy with observations.

Our results confirm that the galaxy clustering observed at small ($\lesssim 10$ Mpc) scales represents the natural product of non–linear gravitational instability for models like CDM and CHDM. But over the length scales considered in this paper, the $S_q$ and VPF statistics do not discriminate between these models, although it appears that the redshift-space VPF may do so on larger scales, and the neighbor analysis of deviations from hierarchical scaling on smaller scales.

# 1 Introduction

In recent years, the clustering properties of galaxies in both simulations and observational data sets have been studied with statistics that go beyond the autocorrelation function. However, the only theoretical model that has been analyzed in any detail with these statistics is cold dark matter (CDM). As has now become well known, standard CDM (Blumenthal et al. 1984; Davis et al. 1985) with linear bias (Bardeen et al. 1986) $b \approx 2.5$ is inconsistent with COBE (Smoot et al. 1992) and other observations of microwave background radiation anisotropies, both angular (e.g., Maddox et al. 1990) and spatial (e.g., Loveday et al. 1992) galaxy correlations, large–scale streaming motions (e.g., Dekel 1992) and cluster correlations (e.g., White et al. 1987; Olivier et al. 1993). On the other hand, COBE–normalized CDM, with $b \approx 1$, is in good agreement with large–scale motions, but has small–scale velocities that are too large (e.g. Davis et al. 1992a and references therein, Klypin et al. 1993, hereafter KHPR93), and various other disagreements with the data. It would be interesting to study the clustering properties of theoretical models that are in better agreement with the data than CDM.

The CHDM spectrum has been shown to agree with the available data at least as well as any other theory that we know about. (For a review of CDM and its main variants, including low–$\Omega$ CDM with a cosmological constant, and tilted CDM, see e.g. Liddle & Lyth 1993a and references therein.) Moreover CHDM is a very well–defined theory with only one additional parameter beyond those of standard CDM, the neutrino mass or $\Omega_\nu$; only the range $\Omega_\nu \approx 0.2$–0.3, corresponding to neutrino mass $m_\nu \approx 4.5 - 7$ eV, is interesting. Basic properties of CHDM (or "mixed dark matter") models were worked out some time ago (Bonometto & Valdarnini 1985; Valdarnini & Bonometto 1985; Achilli, Occhionero, & Scaramella 1985; see also Fang, Li, & Xiang 1984); and the fact that CHDM is a promising model for large scale structure was established by several linear calculations (Holtzman 1989; Schaefer, Shafi, & Stecker 1989; van Dalen & Schaefer 1992; Schaefer & Shafi 1992; Taylor & Rowan–Robinson 1992; Holtzman & Primack 1993; Pogosyan & Starobinsky 1993; Liddle & Lyth 1993b). A simplified nonlinear calculation in a 14 Mpc box for $h = 0.5$ ($h$ is the Hubble parameter in units of 100 km s$^{-1}$ Mpc$^{-1}$; in the following we assume $h = 0.5$) has been done by Davis, Summers, & Schlegel (1992), with the initial neutrino fluctuations set equal to zero. Results from detailed N–body calculations in several simulation boxes (14, 50, and 200 Mpc) have been described by KHPR93, which verified that COBE–normalized CHDM, with linear bias $b \approx 1.5$, is in good agreement with observations, including both small– and large–scale velocities. Higher resolution CHDM simulations have been shown to agree well with IRAS and CfA slice power spectra by Klypin, Nolthenius, and Primack (1993, hereafter KNP), and with CfA group properties by Nolthenius, Klypin, and Primack (1993). In this paper we study the clustering properties of these latter simulations.

The presence of the hot dark matter component in CHDM improves on the agreement with data compared to CDM on both small and large scales. For example, CHDM predicts lower

small–scale velocities than CDM but has more cluster correlations than CDM. This is because the relatively high velocities of the light neutrinos even at late times suppresses their clustering on small scales, while their large free–streaming length decreases the fluctuation power in the cold as well as hot components on intermediate as well as small scales. But this raises the interesting question whether the presence of these new length scales in the CHDM model will be reflected in the clustering properties of the dark matter or galaxies in this model. We will show in this paper, by analyzing CDM and CHDM simulations in parallel, that this does not occur.

As is known, the 2–point correlation function provides only a limited statistical description. Further pieces of information are given by the higher–order irreducible correlation functions $\xi_q(\mathbf{x}_1, ..., \mathbf{x}_q)$. The moments of counts in volumes of size $R$ allow one to extract the cumulants $\bar{\xi}_q(R)$, which are the average of the $q$–point function inside the sampling volume (see Section 3). Since $\bar{\xi}_q$ provides an integral description, it suffers less from statistical noise, at the cost of losing geometrical information.

Since the initial analyses of higher–order functions from angular data, hierarchical scaling of $q$–point functions has been detected. This allows $\xi_q$ to be expressed as a linear combination of products of $q$–1 2–point functions, with suitable coefficients (e.g., Groth & Peebles 1977; Fry & Peebles 1978; Sharp, Bonometto & Lucchin 1984). The hierarchical scaling for $\xi_q$ turns into an analogous scaling for the cumulants:

$$\bar{\xi}_q = S_q \, \bar{\xi}_2^{q-1}, \qquad (1)$$

with coefficients $S_q$ independent of the scale. Here $\bar{\xi}_2$ is the variance of the counts, $q = 3$ is for the skewness (see Coles & Frenk 1991, for a discussion about the relevance of the skewness in large–scale structure studies), while $q = 4$ deals with the kurtosis of the distribution.

Saunders et al. (1991) estimated variance and skewness for the IRAS QDOT redshift sample by using Gaussian–shaped cells. Bouchet et al. (1993) realized a similar analysis by using spherical cells on the 1.2 Jy IRAS sample and extracted signals for the correlation functions up to the fifth order. Fry & Gatzañaga (1993b) analyzed the CfA, SSRS and 1.2 Jy IRAS catalogs and detected skewness and kurtosis. A particular attention was paid to take into account effects of redshift distortions. Bonometto et al. (1993) performed a 3– and 4–point function analysis using neighbor counts in the Perseus–Pisces redshift survey (Haynes & Giovanelli 1986). Gatzañaga (1993) applied the method of cell counts to the angular APM galaxy distribution, claiming the detection of a significant signal up to the ninth order. These analyses converge to indicate that the hierarchical scaling always provides a rather good fit. The reduced skewness is $S_3 \simeq 3$ over a quite large scale range, extending from the non–linear ($\bar{\xi}_2 > 1$) regime to the quasi–linear ($\bar{\xi}_2 \lesssim 1$) regime. The reduced kurtosis does not show substantial scale dependence, with $S_4 \sim 10$.

Hierarchical scaling is predicted by models of strongly non–linear clustering, as BBGKY equations (Davis & Peebles 1977; Fry 1984a; Hamilton 1988). In the mildly non–linear regime,

second–order perturbative approaches to fluctuation evolution also generates a hierarchical sequence of cumulants (e.g., Peebles 1980; Fry 1984b). Still adopting a perturbative approach, Juszkiewicz, Bouchet & Colombi (1993) worked out the dependence of $S_3$ on the cell shape, as well as on the spectral index $n$ for a power–spectrum $P(k) \propto k^n$. Assuming top–hat spheres as sampling volumes, they found

$$S_3 = \frac{34}{7} - (n + 3) \qquad (2)$$

for $-3 \leq n < 1$. Catelan & Moscardini (1993) used a Gaussian window to find the dependence on $n$ of both $S_3$ and $S_4$. Bouchet et al. (1992) studied the dependence of $S_3$ on the density parameter $\Omega$.

In the framework of biased galaxy formation (Kaiser 1984; Politzer & Wise 1984; Bardeen et al. 1986), hierarchical correlations in the weak clustering regime are expected for the distribution of high density peaks for both Gaussian (Jensen & Szalay 1986) and non–Gaussian (Matarrese, Lucchin & Bonometto 1986) background statistics, also independent of the prescription to select peaks (Szalay 1988; Borgani & Bonometto 1990). Fry & Gatzañaga (1993a) have shown that selecting high–density peaks from a hierarchical background does not alter the hierarchical behaviour, at least in the weak correlation regime. Amendola & Borgani (1993) argued that hierarchical scaling is the natural outcome for any non–Gaussian distribution, at the scales of weak coherence (i.e., $\bar{\xi}_2 \ll 1$).

Several authors investigated whether hierarchical scaling is supported by N–body simulations. Bouchet & Hernquist (1992) considered simulations based on CDM, HDM and white–noise initial spectra. They found deviations from hierarchical scaling, whose amount decreases as spectra with larger small–scale power are considered. In particular, they detected a scale dependence of the $S_3$ coefficient when going from the linear to the non–linear regime, which is at variance from what observational data indicate. A similar result was found by Lahav et al. (1992). They ascribed the discrepancy to redshift–space distortions and suggested that $S_3$ and $S_4$ appear more constant in redshift space than in real space. Different conclusions were however reached by other authors, who claimed negligible redshift–space distortions of the hierarchical scaling (see, e.g., Fry & Gatzañaga 1993b). Weinberg & Cole (1992) calculated $S_3$ for scale–free spectra with $n = -1, 0, 1$ starting both with Gaussian and non–Gaussian initial conditions. Assuming the CDM power spectrum, Coles et al. (1993) attempted to use skewness to distinguish between Gaussian and skewed initial conditions. Lucchin et al. (1993) estimated cumulants up to the fifth order at different stages of evolution of scale–free spectra with $-3 \leq n \leq 1$. They found that deviations from hierarchical scaling are significant for spectra with more large–scale power.

Such discrepancies with respect to the hierarchical scaling have been interpreted as effects due to limited statistics (Colombi, Bouchet & Schaeffer 1993), inadequate sampling or boundary problems. Nevertheless the same problems should be present also in observational data sets, which however show hierarchical cumulants over a fairly wide dynamical range. This is one of the points that we will address in this paper.

Other than specifying the sequence of correlation functions, hierarchical scaling also has precise consequences on other statistics, such as the void–probability–function (VPF), which probes geometry rather than clustering. The VPF uses information about correlation functions of all orders (e.g., White 1979; see also Section 3 below) and therefore represents a useful diagnostic to characterize global properties of large–scale texture (e.g., Liddle & Weinberg 1993). Properties of VPF have been studied by Fry (1986) in the framework of the hierarchical model and compared with both N–body results (Fry et al. 1989; Bouchet, Schaeffer & Davis 1991; Bouchet & Hernquist 1992) and observational data (e.g., Alimi, Blanchard & Schaeffer 1990; Maurogordato, Schaeffer & da Costa 1992; Bouchet et al. 1993). The above analyses consistently show that VPF closely follows hierarchical predictions made for both observational data and numerical experiments. In this paper, the analysis of VPF is extended to CHDM simulations and results are compared with those for the CDM model.

This paper is organized as follows: In Section 2 we describe our CDM and CHDM simulations as well as our procedure to identify galaxies in the simulations. In Section 3 we give the necessary theoretical background on the counts–in–cells and neighbors methods we use, and on the VPF. In Section 4 we show how the skewness and kurtosis coefficients $S_3$ and $S_4$ are related to the counts in cells and the counts of neighbors. In Section 5 we report our results and comparisons with data both in real and redshift space. Finally, in Section 6, we give our conclusions.

# 2   The Simulations

## 2.1   Description of the Simulations

In this paper we analyze two CHDM and two CDM simulations. All simulations were done for 100 Mpc boxes. (Here and throughout this paper we consider only $\Omega = 1$ models, and assume $h = 0.5$.) The Particle-Mesh (PM) code with $512^3$ force resolution (i.e., 195 kpc) was used for the simulations (Kates et al 1991). The CDM simulations had $256^3 = 16.8 \times 10^6$ cold particles; the CHDM simulations had an additional $2 \times 256^3$ hot particles, for a total of $50.3 \times 10^6$ particles. The cold particles had masses of $4.2 \times 10^9 M_\odot$ in the CDM simulations and $2.9 \times 10^9 M_\odot$ in the CHDM simulations; the hot particles had masses of $6.3 \times 10^9 M_\odot$ in the CHDM simulations. Each pair of the hot particles was given oppositely directed "thermal" velocities drawn from a redshifted Fermi–Dirac distribution (see KHPR93). The power spectrum of Bardeen et al. (1986, eq. G3) was used to set initial conditions for the CDM simulations. The CHDM simulations were made for the case $\Omega_{cdm} = 0.6$, $\Omega_{baryon} = 0.1$, and $\Omega_\nu = 0.3$, using the power spectra given by KHPR93.

Two CHDM simulations were done, $CHDM_1$ and $CHDM_2$, with different realizations of random numbers. The CDM simulations were done with exactly the same random numbers as in $CHDM_1$, so that all large features in these simulations correspond. The two CHDM simulations were started at redshift $z = 15$ and normalized to linear bias parameter $b = 1.5$, i.e. to the COBE amplitude ($17\mu$ for the rms quadrupole anisotropy). No corrections for a slight non–Zel'dovich tilt or for gravity waves predicted by inflationary models (see, e.g., Liddle & Lyth 1993b; Lucchin, Matarrese & Mollerach 1993; Turner 1993) were applied. For a 100 Mpc box simulation those corrections could be largely compensated by decreasing the baryonic fraction and/or lowering the mass of hot particles.

One of the CDM simulations (CDM1) was started at $z = 27.5$ with bias $b = 1$, which corresponds to COBE normalization; if we regard the starting redshift as being $z = 18$, then the results at zero redshift correspond to the linear bias parameter $b = 1.5$ (CDM1.5). We consider both CDM1 and CDM1.5 as useful comparison simulations: CDM1 is normalized exactly the same as $CHDM_1$ on large scales, and CDM1.5 has the same variance as the CHDM simulations on the 16 Mpc bias–normalization scale.

The details of the simulations are presented in KNP; Nolthenius, Klypin, & Primack (1993) summarizes some of the results. Two anomalies in these simulations should be mentioned. First, the largest waves in the $CHDM_1$ and CDM simulations had amplitude about 1.3–1.4 times larger than expected for an average realization. This was a statistical fluke, but not an unlikely one. We estimate that this much extra large–scale power would be generated roughly once in ten or twenty realizations, and it can also be considered as roughly compensating for the finite size of the simulations. $CHDM_2$ has a more typical power spectrum; comparison of these two CHDM simulations thus gives some idea of the cosmic variance. Second, two errors were initially made in setting up the initial conditions for the simulations: (a) the fit to the cold spectrum was 15% too low at large wavenumbers $k \sim 10$ Mpc$^{-1}$, and (b) the initial velocities of the cold particles were 20% too high at large $k$. These are described in more detail in the Note Added in Proof to KHPR93. These errors were fully corrected for $CHDM_2$ and the results we report here are obtained after such revision. However, as is also outlined in KHPR93, the effects of the two errors almost precisely compensate, as was verified on the basis of a particle–by–particle comparison. The changes on the statistical parameters are henceforth absolutely marginal. Although $CHDM_1$ was not reprocessed after correction, there is no reason to believe that such changes would modify statistical conclusions. Our simulations therefore represent CDM and CHDM sample universes reasonably accurately. But in order to compare them with observational data, we must identify galaxies in the simulations.

## 2.2   Galaxy identification

The problem of identifying luminous galaxies in dissipationless simulations of *dark* matter is a long–standing problem. A generally accepted view is that baryonic material should collect and switch on in high–density regions, although it is not clear how to identify such regions. Different prescriptions have led to rather different conclusions about the clustering properties of the resulting galaxies. A classical prescription is based on the bias paradigm for galaxy formation, which identifies galaxy sites with the peaks of the initial (linear) density field whose height exceeds a fixed threshold (Kaiser 1984; Bardeen et al. 1986). In this approach, one selects in N–body simulations the nearest particles to the peaks of the density field on the grid and follows them during the evolution (e.g., Davis et al. 1985; Park 1990; Valdarnini & Borgani 1991). The initial density field is suitably smoothed over an appropriate scale, and the threshold is chosen to produce the correct amplitude of the 2–point correlation function at the present time (i.e., when $\xi$ takes the observed slope, $\xi(r) \propto r^{-1.8}$).

High resolution simulations made it possible to test this prescription, by identifying the halos of DM particles at the end of the evolution and by checking whether peak particles are associated with such halos. It was found ( Kates et al. 1991, Katz, Quinn & Gelb 1992) that a large fraction of massive halos in the evolved density field contains no peak particles, thus casting serious doubt on the reliability of the original peak–biasing prescription. Gelb & Bertschinger (1993) identified galaxies as massive halos of the evolved field (i.e. as local density maxima). However, for a CDM initial spectrum, they found that on small scales the halos are less correlated than real galaxies. They argued that this is due to overmerging occurring in dissipationless simulations, which merges nearby halos, thereby decreasing the number of close galaxy pairs. To overcome this problem, they devised prescriptions to break up massive halos, so as to recover the merged halos. We use the same idea of breaking "overmergers" to smaller halos, but in detail our procedure is slightly different.

KHPR93 identified galaxies in their CDM and CHDM simulations by referring to the density field assigned on the grid at a given moment. For the purpose of studying the correlation properties of the galaxies, in which we are also interested, they identified those cells where the total density ("cold" plus "hot") exceeds the average density by a fixed amount. In their 50 Mpc PM simulation with $256^3$ resolution and $128^3$ particles (thus, the same spatial and mass resolution as in our 100 Mpc PM simulations), they argued that by taking overdensity 50 one gets a reasonable prescription. Density maxima selected according to this threshold criterion are assigned a weight $w = \rho/\langle \rho \rangle$, where the density is estimated at the position of the maximum. One could consider the weight as sort of "luminosity" weighting.

In the present analyses, we pass from the weight $w_i$ to the *number of galaxies* $N_i$ assigned to the $i$-th maximum by taking approximately equal weights for all *galaxies*. Galaxy identification proceeds as follows. Let us assume that the mean galaxy separation is $d$. Then the total number of galaxies in the simulation box is $N_{gal} = (100\,\mathrm{Mpc}/d)^3$. Each $i$-th density maximum gives $N_i = [w_i/\bar{w}]$ galaxies, where $[x]$ is the integer part of $x$ and $\bar{w}$ is a free parameter ($\approx$ weight per "galaxy"). The parameter $\bar{w}$ is chosen in such a way that the number of "galaxies" $\sum_i N_i$ is equal to the expected number $N_{gal}$ of galaxies inside our box.

This prescription produces a small mean ratio $N_i/w_i$ for peaks near the threshold where, on average, a weight $\simeq 1.5\bar{w}$ is required to give a galaxy. For $w_i >> \bar{w}$, instead, the average weight–per–galaxy is $\simeq \bar{w}$. This what one wants from the algorithm. Most halos were resolved and correspond to one galaxy. Very large "overmergers" (the number of them is very small) are broken into smaller galaxies.

We shall report results obtained choosing two mean separations $d = 5$ Mpc and $d = 9$ Mpc corresponding to usual and very bright galaxies. In Figure 1 we show the distribution of DM particles compared to that of *brighter* galaxies. Both CDM and CHDM$_1$ runs are based on the same assignments of initial phases, so that the emerging structures are directly comparable. However, the effect of selecting galaxies from high–density peaks is more pronounced for CHDM: underdense regions turn out to be nearly devoid of galaxies, while more field population is present in the CDM case. CHDM has better defined filaments, while filaments for CDM appear to be broken. CDM clusters have a rounder shape in real space than those in the CHDM simulations.

In Figure 2, we plot the 2–point correlation function for both galaxy populations (short–dashed and solid lines are for brighter and fainter galaxies, respectively) and compare it with power–law model $\xi(r) = (11/r)^{1.8}$. In Table 1 we give the best fitting parameters of the power–law $\xi(r)$ for all the models, for both galaxies and DM particles (see also KHPR, in particular for the comparison real space *vs.* redshift space). The CDM model at $b = 1.5$ produces too weakly clustered galaxies and evolution needs to proceed until $b = 1$ to reach an adequate level of correlation amplitude. For this more evolved configuration, Table 1 shows the presence of a remarkable antibiasing, the DM particles being characterised by a larger correlation length than galaxies. This fact is essentially due to the presence of a large overmerging, which takes place in the evolution of the CDM spectrum (see also Gelb & Bertschinger 1993); as gravitational clustering goes on from $b = 1.5$ to $b = 1$, small structures merge together to form very massive halos. Therefore, the number of galaxy halo pairs at small separation is suppressed, thus decreasing the corresponding correlation amplitude. Consistently with this picture, no antibiasing is present for the CHDM simulations. In fact, for this spectrum the power is spread over a larger scale range, instead of being concentrated at small scales like for CDM. Therefore, the hierarchical merging of smaller structures into larger ones is less relevant and a substantial amplification of the correlation amplitude is attained by selecting massive halos in the final particle configurations. Note that the CHDM1 simulation exhibits stronger clustering, which is caused by the anomalous large–scale power in that realization. A weaker $\xi(r)$ is produced by the CHDM$_2$ run, which reproduces fairly well the observational $\xi(r)$ over the whole range of sampled scales. Note the effect of taking rarer galaxies, which leads to systematically stronger clustering, a sort of *luminosity biasing*. In all cases, $\xi(r)$ steepen at small ($\lesssim 2$ Mpc) scales. This is probably due to the fact that the galaxy identification procedure assigns many galaxies to very high–density peaks and generates a large number of very close pairs.

# 3 The Statistical Background

## 3.1 Count–in–Cell Analysis

Let us consider a generic density field $\rho(\mathbf{x})$ and suppose we sample it with volume elements of size $R$, whose shape is described by the window function $W_R(\mathbf{x})$. The resulting observable quantity is the smoothed field

$$\rho_R(\mathbf{x}) \;=\; \frac{1}{V_R} \int d^3 y \, \rho(\mathbf{y}) \, W_R(\mathbf{x} - \mathbf{y}) \,, \qquad (3)$$

which is the local average of the density within the sampling volume $V_R = \int d^3 x \, W(\mathbf{x})$. If $p(\rho_R)$ represents the probability density function (PDF) of the continuous variable $\rho_R$, then the statistics of the distribution is fully described by the moment generating function (MGF)

$$M(t) \;=\; \langle \, \exp\left( t \rho_R / \bar{\rho} \right) \rangle \;=\; \int d\rho_R \, p(\rho_R) \, e^{t \rho_R / \bar{\rho}} \,, \qquad (4)$$

or, equivalently, by the cumulant generating function

$$K(t) \;=\; \ln M(t) \;=\; \sum_{q=1}^{\infty} \frac{\bar{\xi}_q(R)}{q!} \, t^q \,. \qquad (5)$$

Here $\bar{\xi}_1(R) \equiv 1$, while for $q \geq 2$ $\bar{\xi}_q(R)$ are the $q$-th order cumulants,

$$\bar{\xi}_q(R) \;=\; \frac{1}{V_R^q} \int d^3 x_1 \dots d^3 x_q W_R(\mathbf{x}_1) \dots W_R(\mathbf{x}_q) \xi_q(\mathbf{x}_1, \dots, \mathbf{x}_q) \,, \qquad (6)$$

which is the average value of the irreducible $q$-point correlation function. From eq. (4), the PDF is expressed as the inverse transform of the MGF as

$$p(\rho_R) \;=\; \frac{1}{2\pi i \, \bar{\rho}} \int_{-i\infty}^{+i\infty} dt \, M(t) \, e^{-it \rho_R / \bar{\rho}} \,. \qquad (7)$$

Instead of dealing with continuous distributions, in the analysis of galaxy catalogues as well as of N-body simulations one considers discrete point distributions. Therefore, a suitable prescription is required in order to relate the statistics of the underlying density field to that of its discrete realization. A usual assumption is that the point distribution one deals with represents a Poissonian realization of an underlying continuous field. Let $\bar{N}$ be the average number of objects within $V_R$. In a random realization of a peculiar value of $\rho_R$, the actual number of points must obviously be an integer. Its expected (non–integer) value is $(\bar{N}/\bar{\rho})\rho_R$, and fluctuations around this value are described by a Poissonian statistics. The PDF for a Poisson process $\varphi$ with mean $\bar{\varphi}$ is $p_P(\varphi) = \sum_{N=0}^{\infty} \frac{\bar{\varphi}^N}{N!} e^{-\bar{\varphi}} \delta_D(\varphi - N)$. The PDF for a process $x = \varphi \bar{\rho} / \bar{N}$ is therefore $p_P(x) = p_P(\varphi) \bar{N} / \bar{\rho}$. Accordingly, the MGF reads

$$M_P(t) \;=\; \int dx \, p_P(x) \, e^{tx/\bar{\rho}} \;=\; \exp\left[ \frac{\bar{N}}{\bar{\rho}} \, \rho_R \left( e^{t/\bar{N}} - 1 \right) \right] \,. \qquad (8)$$

This procedure concerning a particular $\rho_R$ is to be averaged over all the possible realizations of the $\rho_R$ process. In this way we obtain the MGF for the discrete counts, which reads

$$M_{disc}(t) = \int d\rho_R \, p(\rho_R) \exp\left[\frac{\bar{N}}{\bar{\rho}} \rho_R \left(e^{t/\bar{N}} - 1\right)\right] = M\left[\bar{N}\left(e^{t/\bar{N}} - 1\right)\right] . \qquad (9)$$

The discrete nature of the point distribution is therefore accounted for by the change of variable $t \to \bar{N}(e^{t/\bar{N}} - 1)$ in the functional dependence of $M(t)$, which leaves the variable unchanged in the limit $\bar{N} \to \infty$.

As for the PDF, in the discrete case eq.(7) gives

$$p(\rho_R) = \frac{1}{2\pi\bar{\rho}} \int_{-\infty}^{+\infty} dt \, M[\bar{N}(e^{it/\bar{N}} - 1)] \, e^{-it\rho_R/\bar{\rho}} . \qquad (10)$$

Since the variable $e^{it}$ takes values only on the unit circle of the complex plane, the MGF turns out to be a periodic function. Therefore, its Fourier transform can be written as a sum of Dirac $\delta$-functions:

$$p(\rho_R) = \frac{\bar{N}}{\bar{\rho}} \sum_{N=-\infty}^{+\infty} \delta_D\left(\frac{\rho_R}{\bar{\rho}} - \frac{N}{\bar{N}}\right) P_N(R) . \qquad (11)$$

The PDF vanishes except for a discrete set of values of $\rho_R/\bar{\rho}$ as it must for a point distribution. In the above expression, the coefficients $P_N(R)$ are

$$P_N(R) = \frac{1}{2\pi i} \oint dy \, y^{-(N-1)} M[\bar{N}(y - 1)] . \qquad (12)$$

For analytical $M(t)$ all the $P_N$'s for $N < 0$ vanish, so that they acquire the meaning of probabilities of finding $N$ points inside a volume of size $R$. For $N \to \infty$ and $\bar{N} \to \infty$, with fixed $N/\bar{N}$, eq.(11) gives back the continuous limit $P_N(R) = (\bar{\rho}/\bar{N})p(\rho_R)$, with the effective density variable given by $\rho_R/\bar{\rho} = N/\bar{N}$.

The statistics of the point distribution can be described in terms of the central moments $\mu_q = \langle (N - \bar{N})^q \rangle / \bar{N}^q$, where the moments of counts

$$\langle N^q \rangle = \sum_{N=1}^{\infty} P_N \, N^q = \left.\frac{d^q M_{disc}(t)}{dt^q}\right|_{t=0} \qquad (13)$$

are the coefficients of the McLaurin expansion of the discrete MGF. According to the above relations and following the definition (6) of cumulants, it is possible to express $\bar{\xi}_q$, which characterize the underlying continuous field, in terms of the the measured moments of discrete counts. At the lowest orders, it is

$$\mu_2 = \frac{1}{\bar{N}} + \bar{\xi}_2$$

$$\mu_3 = -\frac{2}{\bar{N}^2} + 3\frac{\mu_2}{\bar{N}} + \bar{\xi}_3$$

$$\mu_4 = \frac{6}{\bar{N}^3} - 11\frac{\mu_2}{\bar{N}^2} + \frac{\mu_3}{\bar{N}} + 3\mu_2^2 + \bar{\xi}_4 \,, \tag{14}$$

while more cumbersome relations hold at higher orders. As expected, all the shot–noise corrections vanish for very large $\bar{N}$ values. It is clear that eqs.(14) are based on the assumption that the point distribution represents a random sampling of an underlying continuous field. However, if the particles trace the high–density peaks, they are far from being a Poissonian sampling. For this reason, some care must always be payed in the application of shot–noise corrections, especially when the sampling rate is very low (e.g., Borgani et al. 1993).

In our analysis we sample the simulation box with cubic grids of varying spacing. Therefore, at each scale $r$ there are $(L/r)^3$ non–overlapping sampling volumes.

A further question concerns the scale range where properly testing the development of non–linear clustering with the available simulations. At small scales we are limited by the finite force resolution. Although the nominal resolution should be given by the size of the mesh on which the gravitational potential is resolved, we prefer to adopt a more conservative approach. Therefore, we include $4^3$ pixels in the smallest considered cell. In this way, we are fairly sure that any shot–noise effect is negligible. Limitations at large scales are naturally imposed by periodic boundary conditions, which led to consider cell sizes larger than $L/2$. However, effects of finite simulation volume are present already at smaller scales. For instance, Kauffman & Melott (1992) found in their simulations that self–similarity in the void distribution is broken for void sizes larger than $L/4$. For these reasons, we limit our analysis to the scale range $1/128 \lesssim r/L \lesssim 1/4$. Even with these restrictions, the high resolution of our simulations allows us to span the largest dynamical range considered up to now in this kind of analyses.

## 3.2 Moment of Neighbours Analysis

Instead of counting the objects within volumes centered on arbitrarily chosen points, an alternative determination of correlation functions is provided by the moments of neighbour counts. In this case, sampling volumes are centered on each object, so that the relevant quantities are the conditional probabilities $P_N^{(c)}(R)$ of finding $N$ objects in a sampling volume of size $R$, given that such volume is centered on a sample object. Accordingly, the *conditional* central moments $\mu_q^{(c)} = \langle (N_c - \langle N_c \rangle)^q \rangle / \langle N_c \rangle^q$, are obtained from

$$\langle N_c^q \rangle = \sum_{N=1}^{\infty} P_N^{(c)} N^q \,. \tag{15}$$

The statistics of the system is then described by the conditional cumulants

$$\bar{\xi}_q^{(c)}(R) = \frac{1}{V_R^{q-1}} \int d^3x_1 \dots d^3x_{q-1} W_R(\mathbf{x}_1 - \mathbf{x}_0) \dots W_R(\mathbf{x}_{q-1} - \mathbf{x}_0) \, \xi_q(\mathbf{x}_0, \mathbf{x}_1, \dots, \mathbf{x}_{q-1}) \,, \tag{16}$$

which are the average value of the irreducible correlation function within the sampling volume $V_R$, centered on the object located at $\mathbf{x}_o$ (due to the invariance of correlation functions to translations, $\bar{\xi}_q^{(c)}(R)$ does not depend on $\mathbf{x}_0$). In practical applications, one counts the number of neighbours within a distance $R$ from the chosen object, so that the window function $W_R(\mathbf{x})$ describes a top–hat sphere which encompasses a volume $V_R = 4\pi R^3/3$. For $q = 2$, the variance is expressed in terms of the first–order moment of counts as

$$\bar{\xi}_2^{(c)}(R) \; = \; \frac{\langle\, N_c\,\rangle}{\bar{N}} - 1 \,, \tag{17}$$

where $\bar{N}$ is the expected number of objects within $V_R$.

A formalism analogous to that of the §3.1 can be introduced to relate the continuous statistics, described by $\bar{\xi}_q^{(c)}$, to the discrete one, described by $\mu_q^{(c)}$. However, due to the presence of conditional probabilities, eqs.(14) must be modified into (Peebles 1980)

$$\mu_2^{(c)} = \frac{1}{\langle\, N_c\,\rangle} + C_3 \,;$$

$$\mu_3^{(c)} = -\frac{2}{\langle\, N_c^2\,\rangle} + 3\,\frac{\mu_2^{(c)}}{\langle\, N\,\rangle} + C_4 \,. \tag{18}$$

In the above equations the structure of the discreteness terms is the same as that in eqs.(14), provided that the cumulants are replaced by the more complex quantities

$$C_3 \;=\; \frac{1}{(1 + \bar{\xi}_2^{(c)})^2} \left[\bar{\xi}_3^{(c)} - (\bar{\xi}_2^{(c)})^2 + \frac{1}{V^2}\int d^3\boldsymbol{x}_1\,d^3\boldsymbol{x}_2\,\xi(\boldsymbol{x}_{12})\right] \;;$$

$$C_4 \;=\; \frac{1}{(1 + \bar{\xi}_2^{(c)})^3} \left[\bar{\xi}_4^{(c)} - 3\bar{\xi}_2^{(c)}\bar{\xi}_3^{(c)} + 2(\bar{\xi}_2^{(c)})^3 + \frac{1}{V^3}\int d^3\boldsymbol{x}_1\,d^3\boldsymbol{x}_2\,d^3\boldsymbol{x}_3\xi_3(\boldsymbol{x}_{12},\boldsymbol{x}_{23},\boldsymbol{x}_{23})\right] \;. \tag{19}$$

The two methods based on counts within cells or on neighbour counts should in principle furnish equivalent evaluations of correlation functions. It is however clear that some differences are expected, due to the different way in which the sampling volumes cover the point distribution and the shot–noise terms enter in the relation between continuous and discrete descriptions. In particular, in the neighbour count procedure the number of sampling volumes is the same at all scales, while in the cell count approach the number of volumes decreases as $R^{-3}$. Although this could represent an advantage of the first method, since it provides a better sampling, volumes centered on different points tend to have greater and greater overlaps as $R$ increases. Furthermore, while counting neighbours in a sample of $N$ points represents a computation which scales as $N^2$, the count–in–cell method is much less time consuming. While this does not represent a serious issue for the analysis of the galaxy distribution, which contains a rather limited number of points, it becomes crucial when analyzing the distribution of DM particles (although the analysis carried on here is based on a random subsample of 170,000 DM particles). On the other hand, an advantage of counting neighbours is that, at a fixed correlation order, use is made of smaller powers of counts, which potentially reduces effects of limited statistics.

Since observational data sets have been analyzed with both methods, we will also apply the two procedures in order to make appropriate comparisons between simulations and real galaxy samples. To this purpose, we will show how the moments of cells counts and of neighbour counts are related to correlation functions and give a consistent description of the non–linear clustering.

However, in order to closely compare the outputs of the two analyses, we should bear in mind that a sphere of radius $R$ sample an *effective* volume equal to that of a cubic cell of size $1.6\,R$, quite independent of the slope of the 2–point correlation function (see, e.g., Saunders et al. 1991). Taking this into account, the scale range chosen for the neighbours analysis exactly overlaps with that relevant for the cell–count analysis.

## 3.3  The void probability function

A further useful tool to characterize the statistics of the large–scale galaxy distribution is represented by the void probability function (VPF), which gives the probability of finding no objects within a given randomly placed sampling volume. For $N = 0$, eq.(12) gives the VPF as

$$P_0(R) = M(-\bar{N}) = \exp\left[\sum_{q=1}^{\infty} \frac{(-\bar{N})^q}{q!} \bar{\xi}_q(R)\right] \tag{20}$$

($\bar{\xi}_q(R)$'s are defined in eq.(6); $\bar{\xi}_1(R) \equiv 1$). Since $P_0$ conveys information about correlations of any order, the VPF statistic has been suggested as a useful tool to provide a global clustering characterization. Note, however, that $P_0$ depends only on the number of non–empty cells, with no regards to the number of objects contained inside them. For this reason, it provides only a description of the geometry, rather than of the clustering, of a point distribution. For a completely uncorrelated (i.e. Poissonian) distribution, it is $P_0(R) = \exp\left(-\bar{N}_R\right)$, so that any departure of the quantity

$$\sigma(\bar{N}_R, R) = \frac{-\log(P_0)}{\bar{N}_R} \tag{21}$$

from unity represents the signature for the presence of clustering.

The assumption of hierarchical correlations implies that the coefficients $S_q = \bar{\xi}_q / \bar{\xi}_2^{q-1}$ do not depend upon $R$. (We define $S_2 \equiv S_1 \equiv 1$.) Henceforth, under this hierarchical assumption, it will be convenient to introduce the scaling variable $N_c = \bar{N}_R \bar{\xi}_2(R) \propto R^{3-\gamma}$. Owing to eqs. (20) and (21) it follows that

$$\sigma(N_c) = \sum_{q=1}^{\infty} \frac{(-\bar{N}_c)^{q-1}}{q!} S_q, \tag{22}$$

and the assumption of hierarchical scaling is then expressed by the fact that all scale dependence goes through $N_c$. Therefore, while the value taken by $\sigma - 1$ states the deviation of the distribution from Poisson, in the hierarchical scaling regime, the scale dependence of $\sigma$ can be expressed directly through $N_c$ (and not through $\bar{N}_R$ and $R$ separately).

In the analysis of their scale-invariant model, Balian & Schaeffer (1989) found that for asymptotically large $N_c$, the power-law relation $\sigma(N_c) \propto N_c^{-\omega}$ should hold, with $0 < \omega < 1$. However, in the framework of hierarchical correlation pattern, several models have been proposed, each of which provides a different expression for the VPF. These models will be tested in the following against the outputs of our simulations.

Among these models is the thermodynamical model (e.g., Saslaw & Hamilton 1984) which predicts (Fry 1986)

$$\sigma(N_c) \; = \; (1 + N_c)^{-1/2} \, . \tag{23}$$

A further model has been proposed by Fry (1985) and describes the galaxy clustering as due to a Poissonian distribution of clusters, each containing a suitable amount of members. The resulting hierarchical Poisson distribution gives

$$\sigma(N_c) \; = \; \frac{1 - e^{-N_c}}{N_c} \, . \tag{24}$$

The negative binomial model, originally introduced by Carruthers & Shih (1983), predicts

$$\sigma(N_c) \; = \; \frac{\log{(1 + N_c)}}{N_c} \tag{25}$$

and has been shown by Gatzañaga & Yokoyama (1993) to provide a quite good fit to CfA data.

Finally, the phenomenological model

$$\sigma(N_c) \; = \; \left(1 + \frac{N_c}{2\omega}\right)^{-\omega} \tag{26}$$

has been proposed by Alimi et al. (1990), which found a best fit to the CfA data for $\omega = 0.50 \pm 0.15$ (note that for $\omega = 0.5$ eq.[26] coincides with the thermodynamical model prediction).

In the next section, we will explicitly evaluate the third– and fourth–order cumulants and the VPF for CDM and CHDM initial spectra. Other than verifying the dependence of these quantities on the initial spectrum and on the evolutionary stage, we will also concentrate on their dependence on the population chosen to trace the density field, namely DM particles and galaxies, identified to correspond to high density peaks as explained in §2.2.

# 4   Correlation Functions from Moments of Neighbour and Cell Counts

In this section we shall report the relations existing between moments of counts and correlation functions. We assume a pure power–law shape for the 2–point correlation function, $\xi(r) = (r_o/r)^\gamma$, over all the relevant scales and hierarchical model expressions for 3– and 4–point correlation functions:

$$\xi_3(x_1, .., x_3) \; = \; Q[\xi_{12}\xi_{13} + ... (3 \text{ terms})] \, ; \tag{27}$$

$$\xi_4(x_1,..,x_4) \;=\; Q_a[\xi_{12}\xi_{23}\xi_{34} + ... (12\ \text{terms})] + Q_b[\xi_{12}\xi_{13}\xi_{14} + ... (4\ \text{terms})]\,; \tag{28}$$

According to the definition (6) of cumulants of cell–counts, it is

$$\overline{\xi}_2(R) = J_2 \left(\frac{r_o}{R}\right)^{\gamma}\,; \tag{29}$$

$$\overline{\xi}_3(R) = 3J_3 Q \left(\frac{r_o}{R}\right)^{2\gamma}\,; \tag{30}$$

$$\overline{\xi}_4(R) = \left[12 J_{4a} Q_a^{(4)} + 4 J_{4b} Q_b^{(4)}\right] \left(\frac{r_o}{R}\right)^{3\gamma}\,. \tag{31}$$

The quantities $J_2$, $J_3$, $J_{4a}$, $J_{4b}$ are numerical factors depending on $\gamma$ and on the choice of the window function $W_R$ of Section 3.1. Their explicit expressions are given in the Appendix.

The analysis of count–in–cells is performed to work out the coefficients

$$S_3 \;=\; \overline{\xi}_3(R)/[\overline{\xi}_2(R)]^2\,, \tag{32}$$

$$S_4 \;=\; \overline{\xi}_4(R)/[\overline{\xi}_2(R)]^3\,, \tag{33}$$

(reduced *skewness* and *kurtosis*). Hierarchical scaling prescribes that $S_3$ and $S_4$ are scale independent (and so should be all $S_q$'s). In fact, owing to eqs.(29)–(31), we have

$$S_3 \;=\; \eta\,(\gamma)\,Q\,, \tag{34}$$

$$S_4 \;=\; \zeta(\gamma)\left[Q_a^{(4)} + \tau(\gamma)\,Q_b^{(4)}\right] \;=\; \zeta(\gamma)\,Q^{(4)}\,. \tag{35}$$

The factors $\eta$, $\zeta$, $\tau$ depend on $\gamma$ and on the shape of the window function through the $J_q$ integrals, while

$$Q^{(4)} \;=\; Q_a^{(4)} + \tau(\gamma)\,Q_b^{(4)}\,. \tag{36}$$

The coefficients were evaluated numerically by Montecarlo integration in the case of cubic cells and for $\gamma$ in the range 1.2 to 2.2. Figure 3a shows $\eta$, $\zeta$ and $\tau$ as functions of $\gamma$. With an approximation however better than $\sim 6\%$, they can be assumed constant with values

$$\eta = 3.00 \qquad \zeta = 12.4 \qquad \tau = 0.34 \tag{37}$$

for $1.2 \leq \gamma \leq 2.3$. Therefore, to a good approximation,

$$S_3 \simeq 3Q \qquad ; \qquad S_4 \simeq 12.4\,Q^{(4)}\,. \tag{38}$$

In a similar fashion, the hierarchical cumulants for counts of neighbours become

$$\overline{\xi}_2^{(c)}(R) \;=\; K_1 \left(\frac{r_o}{R}\right)^{\gamma}\,; \tag{39}$$

$$\overline{\xi}_3^{(c)}(R) \;=\; Q \left(K_1^2 + 2K_2\right) \left(\frac{r_o}{R}\right)^{2\gamma}\,; \tag{40}$$

$$\overline{\xi}_4^{(c)}(R) \; = \; \left[ 6 \left( K_1 K_2 + K_{3a} \right) Q_a^{(4)} + \left( K_1^3 + 3 K_{3b} \right) Q_b^{(4)} \right] \left( \frac{r_o}{R} \right)^{3\gamma} \; . \tag{41}$$

Again, $K_1$, $K_2$, $K_{3a}$ and $K_{3b}$ are numerical coefficients depending only on $\gamma$ and on the choice of the window function (see the Appendix for their explicit expressions).

As in the case of count–in–cells, we can define *conditional* skewness and kurtosis ($S_3^{(c)}$ and $S_4^{(c)}$). If hierarchical scaling holds, we derive expressions analogous to eqs.(35), with corresponding coefficients $\eta^{(c)}$, $\zeta^{(c)}$ and $\tau^{(c)}$. Their numerical evaluation in the case of top–hat window function yields the curves plotted in Figure 3b. The values

$$\eta^{(c)} = 2.64 \qquad \zeta^{(c)} = 6.96 \qquad \tau^{(c)} = 0.32 \tag{42}$$

are suitable approximations for $1.2 \; \leq \; \gamma \; \leq \; 2.2$ , with an error of $\sim 6\%$ at worse.

# 5   Results and discussion

Making use of the relations found in §4, we performed an analysis of the CHDM$_1$, CHDM$_2$, CDM1.5, CDM1 simulations, at different evolution degrees, both on the basis of actual galaxy coordinates and considering the apparent distribution in space which would result if distance is estimated through redshifts (*redshift space*); galaxy redshifts are evaluated taking into account peculiar velocities, as provided by simulation outputs. We shall refer to each galaxy set, obtained fixing (i) the simulation, (ii) the value of the average intergalactic spacing $d$, (iii) either real space or redshift space, as a *galaxy sample*. It may also be worth mentioning that our outputs for *redshift* space do not show any significant dependence on the point chosen as *observer*. For the neighbour analysis, this has been verified by considering two observers set at a corner and in the center of the simulation output (the two points have a radically different local environment). For the cell analysis the check was extended to a total of 8 different observer settings.

## 5.1   Moments of counts

In Figure 4 filled triangles indicate the variance–skewness relation and filled squares the variance-kurtosis relation for CHDM$_2$ and CDM1.5 models for both DM particles and for galaxies with $d = 9$ Mpc. The dotted lines are the best fit realized under the assumption of hierarchical scaling. The parameters of the fit are reported in Table 2.

Error bars have been estimated by means of a bootstrap resampling procedure, whose implementation proceeds as follows. Let $B(r)$ be the number of non-empty boxes of side $r$. Then, each bootstrap sample is realized by randomly selecting $B(r)$ times the set of counts, allowing for repetition. The analysis is then repeated for each of these bootstrap samples. Under general conditions (see, e.g., Ling, Frenk & Barrow 1986; Efron & Tibshirani 1991), for a sufficiently large number of resamplings the variance inside the bootstrap ensemble should converge to the true sampling variance of the distribution. We found that after 20 such resamplings the error estimates converge in all the considered cases. Error bars plotted in Figure 4 represent $1\sigma$ bootstrap deviations after 20 bootstrap resamplings. Note, however, that these sampling errors are in general very different from the *cosmic variance*, whose amount can be judged from the difference between the two realizations of the CHDM simulations.

Already from this plot, substantial deviations from the hierarchical pattern can be seen for DM distributions (left panels), especially for the $CHDM_2$ model. This is even more apparent from Figure 5, where we plot the scale-dependence of $S_3$ and $S_4$ (filled triangles and squares, respectively). Note that there is no evidence for a scale–range where hierarchical scaling is detected. This agrees with previous analyses by Bouchet & Hernquist (1992) for intial white-noise, CDM, and HDM spectra. They found a smooth decreasing trend for the $S_3$ parameter when going from the strongly non–linear to the weakly non–linear scales (see also Lahav et al. 1993). Colombi, Bouchet & Schaeffer (1993) interpreted such deviations from the expected scale-invariant behaviour as due to the effect of finite statistics. Since any considered distribution contains a finite number of points, there exist at any scale a cell which is characterized by the maximum count $N_{max}$. Therefore, it turns out that $P_N = 0$ for $N > N_{max}$ and the sum in eq.(13) will be truncated, leading to an underestimate of $\langle N^q \rangle$. It is clear that this effect becomes more important as one considers moments of increasing order, which give progressively more weight to richer and richer clusters. Furthermore, since the number counts inside a single cluster approximately scale as $N \propto r^{3-\gamma}$ (here $\gamma$ is the 2–point function slope), the maximum allowed value for the variable $N/\bar{N}$ decreases by increasing the size of the sampling volume. Therefore, finite sample effects become more relevant not only considering higher order moments, but also going to larger scales. This induces a spurious scale–dependence of the $S_q$ coefficients, which decrease even for intrinsically hierarchical distributions. Colombi et al. (1993) checked this effect in detail for their CDM simulations. After suitably correcting for finite statistics, they found a much better agreement with the hierarchical scaling. Since any spectra with more large–scale power produces coherent structures of increasing size, a greater number of points is required to adequately sample it. As a consequence, we expect in our case that effects of finite statistics are more apparent for the CHDM model than for the CDM one. This is in fact confirmed by the plots in Figure 5 where we detect larger deviations from hierarchical scaling for CHDM than for CDM, the second providing systematically smaller $S_q$ values, with a shallower scale–dependence. Significant differences appear even between the two CHDM realizations, the coefficients for the second run being significantly smaller than those for the first run. It is also interesting to note that the CDM model develops a higher degree of hierarchical scaling, especially at a later evolutionary stage. In fact, as the clustering evolves from $b = 1.5$ to $b = 1$, the non-linearity scale shifts to larger values and the $S_3$ profile becomes flatter.

Although such departures from hierarchical scaling could also not be intrinsic to non–linear

DM clustering, nevertheless it is at variance with observational evidence, which points toward $S_3 \simeq$ const for the galaxy distribution over a quite large range of scales. Lahav et al. (1993) argued that the difference between observations and numerical simulations could partly be due to redshift space distortions. To check this, we plot in Figure 5 the $S_3$ and $S_4$ coefficients for all the considered models, from both real space (filled symbols) and redshift space (open symbols) analysis. We confirm that redshift-space distortions tend systematically to flatten the $S_q$ profile, although they act in different ways, according to the amount of large-scale power in the initial spectrum. In fact, differently from all the other cases, in the CHDM$_1$ model the hierarchical coefficients are decreased in redshift space. Therefore, not only the $S_q$ values, but also their dependence on finger-of-God distortions, are sensitive to the choice of the initial spectrum. However, such results from numerical simulations are at variance with respect to indications coming both from other numerical experiments and from observational data sets. Coles et al. (1993) analyzed CDM simulations obtained with both Gaussian and non–Gaussian initial conditions and found that no significant variations of the variance–skewness relation are induced by redshift–space distortions. Fry & Gatzañaga (1993b) analyzed different galaxy redshift samples, with particular care to account for redshift distortions. They found that, although the moments of counts are significantly affected by redshift–space distortions, the quantities $S_3$ and $S_4$ are not, with typical values $S_3 \simeq 2$ and $S_4 \simeq 6$, also quite independent of the considered sample. This result is remarkably different from those obtained from our analysis of CDM and CHDM distributions, as well as from the analyses by Bouchet & Hernquist (1992) and Lahav et al. (1992) for several cosmological spectra, which gives much larger values for such coefficients.

In order to understand the origin of these discrepancies between numerical and observational data, we analyze the distribution of galaxies, identified from high peaks of the evolved density field according to the prescription outlined in Section 2. In the right panels of Figure 4 we report the results for the "bright" galaxies. The major effects of selecting peaks is to enlarge the range of values taken by the $\bar{\xi}_2$ variance by several orders of magnitude, and to remarkably improve the fit to the hierarchical scaling. This is even more apparent from Figure 6, where we plot the reduced skewness $S_3$ as a function of the scale for the two CHDM runs (Fig. 6a) and for the CDM model at $b = 1.5$ and $b = 1$ (Fig. 6b). The effect of taking high peaks is really dramatic at the scales of non–linear clustering. In all the considered cases the galaxy distribution produces a well defined hierarchical scaling over the whole range of non–linearity scales. This finding seems rather surprising since effects of finite sampling should be even more important for the galaxy distribution, which has a much smaller average density that the DM particles. However, since galaxies are identified to correspond to high–density peaks, the value of the variable $N/\bar{N}$ inside clusters is much larger for galaxies than for DM particles. Therefore, the contributions from counts larger than $N_{max}/\bar{N}$ in eq.(13) becomes less important. This could well be the reason why the analysis of galaxy samples reveals a well defined hierarchical scaling, while the DM distribution from N–body simulations do not.

Only as the linear regime is approached, does the $S_3$ value for both dark matter and galaxies approach the same value. This seems to agree with the fact that perturbation theory, which applies to describe the matter distribution in the mildly non–linear regime, gives $S_3$ values which are reasonably close to those detected for galaxies in the same regime. However, as already observed, effects of finite statistics are expected to play a significant role at the largest considered scales, especially for the DM distribution. Therefore, it is difficult to say whether the decreasing trend of $S_3$ is the imprint of the transition from strongly non–linear to quasi–linear regime, or merely represents a spurious effect of finite sampling.

The values of the $S_3$ coefficient for galaxies is much smaller than for the DM particles and show a decreasing trend as the mean galaxy separation decreases. This can be understood in the framework of linear biasing for galaxy clustering. In fact, density contrast in galaxy number counts and in the DM distribution are related by $\delta_g = b\delta_{DM}$. Therefore, it is $\bar{\xi}_{q,g} = b^q \bar{\xi}_{q,DM}$ for the corresponding cumulants. In the framework of hierarchical scaling, this turns into $S_{q,g} = b^{2-q} S_{q,DM}$ and the coefficients decrease by increasing the biasing, that is by taking more rare peaks.

Furthermore, no significant redshift–space distortions are detected, unlike for the DM distributions, but in accordance with the analysis of Fry & Gatzañaga (1993b) of CfA (Huchra et al. 1983), SSRS (Da Costa et al. 1991) and Strauss et al. (1992) IRAS samples. For CDM models, which generate higher velocity dispersions, there appears some trend for $S_3$ to increase in redshift–space as linearity scales are approached. Simulations reaching scales where $S_q$'s (for $q > 2$) tend to vanish in real space would be however valuable to deepen our understanding of the transition from linear to non–linear regimes. Velocity induced distortions of $S_3$ at these scales, if confirmed, can actually be different for different models and used as a test to distinguish amongst them. This comparison could be also based on analytical analyses of large scale quasi–linear dynamical evolution, taking also suitably into account effects due to Poissonian noise.

Within the framework of the simulations considered here, the most striking result, which appears from Figure 6, is that galaxies with $d = 9$ Mpc have $S_3 \simeq 2.5$, quite independent of the model. A similar result also holds for the reduced kurtosis $S_4$, which is plotted in Figure 7 for real–space analysis. Also in this case, selecting galaxies as high–density peaks flattens the $S_4$ coefficient to a much smaller value with respect to that relevant for the DM distribution, quite independent of the initial spectrum. On one hand, these results indicate that, although reduced skewness and kurtosis for the DM clustering resemble memory of initial conditions for structure formation, the same is not true for the clustering of high–density peaks. On the other hand, our analysis shows that both the hierarchical behaviour and the $S_q$ values detected for the galaxy distribution represent the natural outputs of non–linear gravitational clustering coupled with the occurrence of galaxy formation in high–density regions.

In Table 3 we compare our best–fit $S_3$ and $S_4$ values for "bright" galaxies with those obtained from the analysis of galaxy samples (note that Bouchet et al. (1993) found a value $S_3 \simeq 1.5$ for the Strauss et al. (1992) IRAS sample, which is smaller than any other detection; they suggest that this value is an underestimate of $S_3$ for optical galaxies, since IRAS galaxies tend to avoid rich clusters). The values found by Gatzañaga (1993) for the APM angular galaxy sample are larger than any other observational value, especially for the reduced kurtosis. This could be partly due to the increasing trend of the $S_q$ coefficients when the mean galaxy separation decreases (deep angular samples include fainter galaxies than in redshift samples), while also the different analysis technique could play a role. In general, the agreement between results coming from real galaxy samples and our "bright" galaxy distribution is rather good. Although agreement with observational results is in general achieved by all the models within $\lesssim 2\sigma$ level, nevertheless some of these seem to perform better. For instance, as for CDM, the $b = 1.5$ case generates marginally larger $S_3$ values. The same is also true for the CHDM$_1$ realization, which probably represents a configuration with an anomalous excess of large–scale power. Both unbiased CDM and CHDM$_2$, which has a more typical power, are better. Over the scales inspected here no significant variation is found when passing from real to redshift space: velocities seem to give just a minor contribution to estimates of non–Gaussian behaviour.

In Figure 8 we show the effect of dynamical evolution on the high–peak clustering for the two runs of CHDM. In these plots, squares refer to the present time, as also plotted in Figure 6a, while triangles are for the redshift $z = 1$. For both CHDM realizations, the effect of evolution is that of decreasing $S_3$, while approaching the hierarchical scaling. However, some differences exist between the two runs as well as between differently biased galaxy populations. The CHDM$_1$ model shows less clustering evolution. Since this realization has a greater amount of large–scale power, galaxies formed with greater probability in the crests of the long wavelength modes, so that their clustering attains the regime of dynamical stability earlier. This is also the reason for the slower evolution of brighter galaxies, which are identified with relatively higher density peaks. The effect of the evolution on the CDM model can be judged from Figure 6b, by comparing the results for $b = 1.5$ and $b = 1$. Also in this case, the tendency of evolution to produce a better defined scaling with lower hierarchical coefficients is confirmed.

By virtue of the relations outlined in §4, results from counts in cells can be directly compared with those derived from counts of neighbours. In the latter case errors were estimated by essentially the same technique used by Bonometto et al. (1993) in the analysis of the Perseus Pisces redshift sample (PPS). Ten subsamples made of half the galaxies of the original samples were randomly selected. Galaxies of such subsamples were used as centers, while all galaxies were taken as possible companions. All quantities were measured 10 times in this way. Their distribution turned out to be consistent with Gaussian, and this allows to estimate the standard deviation of each measure.

In Figure 9 we plot the values of $Q$ and $Q^{(4)}$, for different scales $R$, as obtained from cells (circles) and neighbours (triangles), both for real and redshift space. Here error bars correspond to 3-$\sigma$'s. In the same plots we also show values and error bars obtained from the analysis of PPS. Such an analysis was performed after some corrections for local motion, Virgo infall and virial fingers. This tentative passage from redshift space data to real space ones can be however much questioned and we hope to be able soon to make a direct comparison among redshift space outputs (both for simulations and real samples).

Errors from counts of neighbours are significantly smaller than those for cells; in a few cases, namely at great scales, the difference is dramatic. This is related to the decrease of the number of volumes, over which the averaging procedure is performed, as the side of the cell increases. It is however important to mention that, within error bars, no significant discrepancy appears between cell and neighbour outputs. The only exception is a point at 6 Mpc for CDM. At such $R$ an irregularity is also visible in the scale dependence of $S_3$, $e.g.$ in Figure 7, and might be a spurious feature.

All the galaxy samples show a general agreement between the measures of $Q$ and $Q^{(4)}$ in real space and the corresponding ones in redshift space, except perhaps for those at the smallest scale considered ($r = 1$ Mpc). However the latter ones should be taken with some caution because such a radius is only five times the resolution of the numerical simulation ($\simeq 0.2$ Mpc).

We are allowed to conclude that all the models considered show a basic hierarchical character of the *galaxy* distributions both in real and in redshift space, independently of the nature of the leading dynamical component (cold or mixed dark matter). The fact that hierarchical scaling also pertains to redshift space distributions proves that peculiar velocity distortions, though significantly affecting the apparent *amount* of clustering (i.e., the moments of counts), operate instead in a mild fashion on the *nature* of the clustering. This conclusion is reinforced by the fact that strong deviations are not present even in the case of unbiased CDM, where peculiar velocities are remarkably higher than in the other models. Their disruptive effects on clustering in redshift space dramatically appear to the eye from Figure 10, where a comparison is made between real and redshift space for a slice of $CHDM_1$ (Fig. 10a) and CDM with b=1.5 (Fig. 10b) and b=1 (Fig. 10c). Here we plot the positions of the galaxies with $d = 5$ Mpc. We do not consider the $CHDM_2$ model, whose different assignment of the initial phases makes difficult any comparison of the distortions of the clustering pattern in redshift space. Although qualitative, the differences shown in Figure 10 are rather significant and make even more remarkable the stability of the hierarchical coefficients against redshift distortions.

Note that, making use of neighbour counts and thanks to their smaller errors especially at large scales, slight deviations from a general hierarchical pattern can be appreciated. In particular, some evidence exists that both $Q$ and $Q^{(4)}$ tend to increase towards smaller scales, namely for the $CHDM_1$ simulation.

This behaviour is present for PPS observational outputs in still a more marked form. Error bars for $CHDM_1$ and PPS almost overlap in the case of the 3-point function, although, in the 4-point case, the PPS trend is stronger than the simulations. In Figure 11 we compare the behaviours of $CHDM_1$ (full circles), $CHDM_2$ (open circles) and PPS (triangles) in more detail. Differences between the two CHDM runs are less significant than those between $CHDM_1$ and PPS. However, in the case of $Q$, PPS behaviour can be probably accounted for by cosmic variance (this is more problematic for $Q^{(4)}$).

A word of caution, however: the analysis of PPS relies on corrections to virial fingers which apply to groups and might have modified some inner characteristics. This reflects a more general problem, as we should remember that our analysis involves mostly scales of non–linear clustering, while analysis of real galaxy samples usually does not separate the non–linear regime from the quasi–linear one. Agreement between theory and observations should be better sought either by analysing PPS and simulations directly in redshift space or passing through the realization of mock galaxy samples, having the same size and selection effects as real samples. This procedure would be linked to a number of assumptions and parameter choices (*e.g.*, on the galaxy luminosity function) that the present approach avoids.

## 5.2   Results on VPF

The VPF has long been recognized to be a useful tool to characterize global properties of the large–scale texture of the galaxy distribution. It is relatively easy to implement and provides statistical information which is in some sense complementary to that of correlation functions, the latter being sensitive to the clustering inside overdensities. The presence of big voids in the galaxy distribution has been suggested to be the signature of biased galaxy formation (e.g., Betancort–Rijo 1990; Einasto et al. 1991). The availability of extended observational data about the VPF from the CfA survey (Vogeley, Geller & Huchra 1991) allowed Liddle & Weinberg (1993) to apply the void statistics to N–body simulations to test different schemes for biased galaxy formation. Weinberg & Cole (1992) used the VPF to distinguish between Gaussian and non–Gaussian initial conditions in their simulations.

As previously shown, the scaling of the quantity $\sigma(\bar{N}, r)$, defined by eq.(21) as the deviation from Poissonian VPF, can be used as a test for the hierarchical ansatz. Fry et al. (1989) applied this statistic to both observational data and simulations and found that the hierarchical scaling gives in general a good description. Bouchet, Schaeffer & Davis (1991) found in their count–in–cell analysis of CDM simulations that the asymptotic scaling $\sigma(N_c) \propto N_c^{-\omega}$ is obeyed by DM particles with $\omega \simeq 0.4$, smaller than that observed from the analysis of galaxy redshift samples (Alimi et al. 1990; Maurogordato et al. 1991). A similar analysis has also been done by Bouchet & Hernquist (1992) for different initial spectra. Again, although the asymptotic behaviour is quite well reproduced, the scaling parameter turns out to be smaller than observed, but with values increasing with the amount of small–scale power in the initial fluctuation spectrum.

Here, we analyze the $\sigma(N_c)$ function for our simulated distribution of bright galaxies with a twofold aim: first, compare with the hierarchical VPF models introduced in §3.3; second, compare with analogous results from observations. In Figure 12 we plot $\sigma(N_c)$ for our brighter galaxies both in real (Fig. 12a) and in redshift (Fig. 12b) space.

As for the error estimate in the VPF analysis, an analytical expression for the standard deviation in $\sigma(N_c)$ has been proposed by Maurogordato & Lachieze–Rey (1987). Under the assumption that counts inside different sampling volumes are independent, these authors found

$$\Delta\sigma \;=\; \frac{1}{\bar{N}}\,\left(\frac{1 - P_0}{N_0}\right)^{1/2}\,, \tag{43}$$

where $N_0$ is the number of empty volumes. It is however clear that, since significant correlations exist over all the considered scales, the assumption of independent counts in different volumes is not satisfied. For this reason, we decided to estimate $\delta\sigma$ through the bootstrap method; each bootstrap resampling at the scale $r$ is obtained by randomly selecting with repetition $(L/r)^3$ times both empty and occupied cells. The resulting $\Delta\sigma$ is that plotted in Figure 12 and turns out to be systematically larger than that provided by eq.(43) by about 50%. In a similar fashion, the plotted errorbars in the $N_c$ variable come from the bootstrap method already described in §5.1.

It is apparent that the phenomenological model, suggested by Alimi et al. (1990; see eq.[26]), always provides a good fit. Even this model, however, is not able to reproduce the data at large scales (corresponding to large $N_c$ values). This fact has been already observed by Bouchet et al. (1991). They argued that, differently from what happens in the overdense regions, in the underdensities the structure of the lattice, on which intial conditions are settled, is essentially preserved. Therefore, on the larger scales of the typical interparticle distance in the void regions gravity has still had no time to build up the hierarchical scaling. For this reason, we perform the best-fit to eq.(26) only up to the $N_c$ values where $\sigma$ starts falling off. In general, a good scaling is observed up to $N_c \sim 10$. Note also that for models with less power at small scales the dynamical evolution inside underdensities proceeds slower. This is the reason for the larger deviations observed for the CHDM runs compared to the CDM ones.

The VPF is decreased in redshift space at large scales, which increases $\sigma$ while leaving it substantially unchanged at small scales (see also Liddle & Weinberg 1993). The only exception is represented by the CHDM$_1$ model, for which the effects of redshift distortion are less relevant (see Figure 12b). The net effect is to partially compensate the spurious fall-off of $\sigma(N_c)$, so to apparently restore to some extent the hierarchical scaling.

In Table 4 we report the best-fit values of the scaling parameter $\omega$ obtained for our simulations in real and redshift space and for observational data sets, which are realized in redshift space. As far as it can be judged from published results, the uncertainties in $\omega$ values for observational data sets come from unweighted least square fitting, so that we also follow this procedure. However, the presence of quite large errorbars, especially in the $N_c$ values, suggests that relative uncertainties as large as $\sim 20\%$ should be expected in the determination of $\omega$. Note that a quite good agreement is always found, except for the analysis by Bouchet et al. (1993) of the IRAS sample. The same authors, however, ascribe the anomalous large $\omega$ value to the fact the IRAS galaxy distribution does not sample the regime where $\sigma$ takes the asymptotic behaviour.

The overall emerging picture for the VPF analysis of our simulated galaxy distributions is that it follows quite closely the hierarchical scaling. Therefore, the hierarchical ansatz does not only provide a good representation of the non–linear clustering, as shown by the high–order correlation analysis, but also of the geometry of the large–scale structure, as described by the VPF. Even in this case, in the region where the scaling of $\sigma(N_c)$ is well detected, significant differences do not occur when passing from real to redshift space or considering different initial spectra. The value of the scaling parameter $\omega$ is always very close to 0.5, thus supporting the reliability of the Saslaw's thermodynamical model of eq.(23) to describe the non–linear clustering of high–density peaks. Vice versa, the large–scale behaviour of the VPF seems to be more sensitive to the spectrum shape, which makes it a potentially useful statistic to discriminate between different models as larger simulations are analyzed.

# 6    Conclusions

In this paper we analysed four simulation made with resolution $\sim 195.3\,\mathrm{kpc}$ in a cubic box of $100\,\mathrm{Mpc}$ a side. This allows us to reach scales not much above the size of individual galaxies, simultaneously inspecting large scale structures up to a range where non–linearity effects begin to weaken their influence. The outputs described two different realizations of CHDM with bias parameter $b = 1.5$ (CHDM$_1$ and CHDM$_2$, respectively) and CDM with $b = 1.5$ and $b = 1$. Of these, the CHDM and CDM1 are normalized consistently with the COBE detection of cosmic background anisotropies, while CDM1.5 has less power but the same bias as the COBE-normalized CHDM runs. The initial density fluctuation spectrum is pure Zel'dovich for all cases.

We evaluated the 3– and 4–point correlation functions and the VPF (void probability function) for the above simulations; $q$–point functions were worked out from the moments of cell and neighbour counts. The possibility that such measurements can show differences between pure CDM and CHDM models was inspected, and only marginal signals indicating a discrimination are detected. A comparison with the outputs of a neighbour analysis carried on by Bonometto et al. (1993) on the Perseus Pisces Redshift Survey (PPS) was also made, although such analysis had been performed after a number of corrections (virial finger compression, virgocentric infall motion cancellation, etc.) which made the comparison not fully homogeneous. Our main conclusions, however, refer to the connection between the observed clustering structure and the expected hierarchical clustering pattern.

It is worth outlining that cell and neighbour approaches have different advantages, the former being much less computationally expensive, while providing in most cases reliable statistical estimates. However, it provides the same sampling rate in the overdense and underdense parts of the distribution. Furthermore, as large scales are considered, only few sampling volumes are allowed for count–in–cells. Bootstrap errors therefore rapidly increase and signal detection is hard. This has been tested also in our analysis. The neighbour approach instead samples regions centered on existing objects and therefore avoids empty regions. The number of counts from which moments are evaluated is set by the number of *galaxies*. Each count however is an *integral* measure, carried on a volume which intersects volumes centered on nearby galaxies. For this reason, some doubt could be raised about the statistical independence of the information provided by overlapping volumes. Within this context errors are evaluated by selecting half of the centers – several times and in a random way – and verifying that different selections lead to results distributed in a Gaussian way. The neighbour method does not become less efficient when greater scales are considered, as the number of counts averaged does not depend on the scale.

By using such techniques, the observed behaviours of DM particles and of *galaxies* were inspected. Substantial deviations of DM particle distributions from the hierarchical clustering pattern were found. No scale range where their behaviour is hierarchical is detected. Also according to previous analyses by Bouchet & Hernquist (1992), Lahav et al. (1993), and Colombi et al (1993), this can be tentatively interpreted as an effect of scarce statistics for DM particles. If this is correct, we should expect that models with greater power over large scales show a stronger deviation from the hierarchical pattern. As a matter of fact, CHDM particles show a stronger deviation from hierarchical clustering than CDM particles, while – between the two different CHDM realizations – $CHDM_1$ indicates an even stronger deviation, still in agreement with the greater large scale power that characterizes this case.

Even though such departures from hierarchical scaling may not be intrinsic to DM non–linear clustering, still it contrasts with observational evidence on the galaxy distribution. Furthermore, the scaling of $S_3$ and $S_4$ coefficients for DM particles is significantly affected by redshift distortion, at variance with indications from observational data analyses (Fry & Gatzañaga 1993).

However the main issue here is that, although the DM particle distribution is non–hierarchical, galaxies are distributed in an almost hierarchical fashion. Galaxies are selected in accordance with the peak distribution of the density field. Selecting peaks produces hierarchical scaling over the whole range of non–linearity. Only when linearity is approached are skewness and kurtosis seen to converge on similar values both for DM and galaxies.

Quite differently from what is noticed for DM, the passage from real to redshift space does not modify the values detected for the hierarchical coefficients, so that they can be efficiently estimated also in redshift space. The values of such coefficients, for the simulations examined here, are reported in Table 2 and turn out to remarkably agree with those detected for observational data sets (Table 3).

The effect of dynamical evolution on clustering was also studied. A comparison with distributions at a redshift $z = 1$ indicates that clustering evolution is smaller for $CHDM_1$ than for $CHDM_2$. The former realization has more power over large scales; therefore galaxies could form earlier on the crests of long wave–length components and attain dynamical stability at earlier times. As gravitational evolution goes on, skewness and kurtosis values decreases until a stable clustering is attained. This is more pronounced for our "fainter" galaxies, which are identified to correspond to lower peaks, thus taking longer to reach a stable clustering.

Results from cells and neighbours can be compared only for galaxies; a neighbour analysis for DM points would be highly time consuming, as the time increases with the number of points squared, for this kind of analysis. In this paper we also provide a set of numerical results allowing a detailed conversion from cells to neighbours. As far as galaxies are concerned, there is a generic agreement between results coming from the two methods, as should be expected. Within the neighbour approach errors do not increase because of lack of sampling, when going towards larger scales. Mostly because of this more precise large scale detection of the hierarchical coefficient, slight deviations from a purely hierarchical behaviour seem to be detectable. Such deviations are within the error bars of the cell approach, but are hidden by their amplitude.

These deviations essentially amount to an increase of hierarchical coefficients towards smaller scales. The trend is almost negligible for CDM with $b = 1$, while it is visible for both CHDM models, although it is even more evident for $CHDM_1$. A similar behaviour was detected with a similar technique in the analysis of PPS. The trend is quite visible both in real and in redshift space.

Neighbour analysis on real samples was carried on for PPS only, up to now. If the trend found there is confirmed by an analysis based on PPS without corrections and on other samples, there is some evidence that this is more easily reproducible with CHDM models than with pure CDM. A comparison between CDM outputs with $b = 1.5$ and $b = 1$ shows that this trend might be linked to the bias level, rather than to the nature of DM. However, as COBE CBR fluctuation detection does not allow standard CDM with $b = 1.5$, this point favours CHDM.

Results on VPF are expected to provide information complementary to correlation functions, being sensible to the geometry of the distribution, rather than on its clustering. Big voids are likely to be a signature of *bias*. It should be however noted that a simulation covering a box whose side is $100 \, \mathrm{Mpc}$ is inadequate to test the scales where big voids are actually present. An inspection of Figure 12 also shows that we can expect a relevant difference between CDM and CHDM, in the passage from real to redshift spaces, as far as the statistical character of voids is concerned. In order to test such effects on a statistical basis we need to sample regions whose size is in the $30$–$50 \, \mathrm{Mpc}$ range. The analysis carried on here is necessarily restricted to smaller scales; here the aim was to compare the VPF with the prediction from different models of hierarchical clustering and with observational data over these scales. It is clear that the phenomenological model of Alimi et al. (1990) always provides a satisfactory fit. However, the allowed value intervals for the parameter $\omega$ always contain $\omega = 0.5$. This gives back the so–called thermodynamical clustering model of Saslaw & Hamilton (1984), still within the framework of hierarchical scaling.

Altogether our results confirm a hierarchical scaling picture for galaxies, in contrast with DM particles. It is notable that the very significant deviations observed for DM do not have dynamical consequences for the galaxy distribution. Here slight discrepancies from a purely hierarchical pattern have quite a different nature and seem to be related to the bias level. The general conclusion is that the hierarchical scaling detected for galaxies at small scales ($\lesssim 10$ Mpc) of non–linear clustering is the natural outcome of non–linear gravitational clustering.

# Acknowledgments

JRP acknowledges support from NSF grant PHY-9024920 and from the University of California, Santa Cruz.

# Appendix

In this appendix we shall provide formal expressions for the coefficients $J_2$, $J_3$, $J_{4a}$, $J_{4b}$ and $K_1$, $K_2$, $K_{3a}$, $K_{3b}$. Suitable combinations of these coefficients yield the quantities $\eta$, $\zeta$, $\tau$ and $\eta^{(c)}$, $\zeta^{(c)}$, $\tau^{(c)}$ of Section 4.

In the expressions of the above coefficients, cubic cells and top–hat spheres are taken for the cell–count and neighbor–count methods, respectively. The side of cubes as well as the radius of spheres is unity. Let also be $v = \int d^3\mathbf{x}\, W_1(\mathbf{x})$ (in the case of cubes $v = 1$, in the case of spheres $v = 4\pi/3$).

The following expressions hold:

$$J_2 = v^{-2} \int d^3x_1\, d^3x_2\, W_1(\mathbf{x}_1) W_1(\mathbf{x}_2)\, x_{12}^{-\gamma} \; ;$$

$$J_3 = v^{-3} \int d^3x_1\, d^3x_2\, d^3x_3\, W_1(\mathbf{x}_1) W_1(\mathbf{x}_2) W_1(\mathbf{x}_3)\, (x_{12} x_{23})^{-\gamma} \; ;$$

$$J_{4a} = v^{-4} \int d^3x_1\, d^3x_2\, d^3x_3\, d^3x_4\, W_1(\mathbf{x}_1) \dots W_1(\mathbf{x}_4)\, (x_{12} x_{23} x_{34})^{-\gamma} \; ;$$

$$J_{4b} = v^{-4} \int d^3x_1\, d^3x_2\, d^3x_3\, d^3x_4\, W_1(\mathbf{x}_1) \dots W_1(\mathbf{x}_4)\, (x_{12} x_{13} x_{14})^{-\gamma} \; ; \qquad (A1)$$

and

$$K_1 = v^{-1} \int d^3x_1\, W_1(\mathbf{x}_1)\, x_1^{-\gamma} \; , ;$$

$$K_2 = v^{-2} \int d^3x_1\, d^3x_2\, W_1(\mathbf{x}_1) W_1(\mathbf{x}_2)\, (x_1 x_{12})^{-\gamma} \; ;$$

$$K_{3a} = v^{-3} \int d^3x_1\, d^3x_2\, d^3x_3\, W_1(\mathbf{x}_1) W_1(\mathbf{x}_2) W_1(\mathbf{x}_3)\, (x_1 x_{12} x_{23})^{-\gamma} \; ;$$

$$K_{3b} = v^{-3} \int d^3x_1\, d^3x_2\, d^3x_3\, W_1(\mathbf{x}_1) W_1(\mathbf{x}_2) W_1(\mathbf{x}_3)\, (x_1 x_{12} x_{13})^{-\gamma} \; . \qquad (A2)$$

In the case of spheres, some integrals can be worked out analytically (see Peebles 1980 and Gaztañaga & Yokoyama 1993). It turns out that

$$K_1 = \frac{3}{\gamma_3} \quad , \qquad J_2 = \frac{9 \cdot 2^{\gamma_3}}{\gamma_3 \gamma_4 \gamma_6} \, ,$$

$$J_3 = \frac{27 \cdot 4^{\gamma_3}}{\gamma_2^2 \gamma_4^2 \gamma_6^2} \left[ \frac{24\, \gamma_{23/24} + \gamma^2\, \gamma_{17/2}}{4\, \gamma_{7/2}\, \gamma_{9/2}} - \frac{4^\gamma \sqrt{\pi}\, \gamma_2\, \Gamma(\gamma_5)}{256\, \Gamma(\gamma_{9/2})} - \frac{4^\gamma \sqrt{\pi}\, \Gamma(\gamma_5)}{256\, \Gamma(\gamma_{11/2})} \right] \; .$$

with $\gamma_s = s - \gamma$.

The relations between such integrals and the coefficients $\eta$, $\zeta$, $\tau$ read

$$\eta = 3 \frac{J_3^{\,2}}{J_2} \; , \qquad \zeta = 12 \frac{J_{4a}}{J_2^3} \; , \qquad \tau = \frac{J_{4b}}{3 J_{4a}}$$

and

$$\eta^{(c)} = 1 + \frac{2K_2}{K_1^2} \;, \qquad \zeta^{(c)} = \frac{6\left(K_1 K_2 + K_{3a}\right)}{K_1^{\,3}} \;, \qquad \tau^{(c)} = \frac{K_1^{\,3} + 3K_{3b}}{6(K_1 K_2 + K_{3a})} \;.$$

Through these relations the curves plotted in Figure 3 were obtained.

# Figure captions

**Figure 1.** Distribution of DM particles compared to galaxies at distance $d = 9\,\mathrm{Mpc}$ ($H_0 = 50$ km s$^{-1}$ Mpc$^{-1}$). CDM and CHDM$_1$ runs are shown.

**Figure 2.** 2–point correlation functions for galaxies (dashed and solid lines correspond $d = 9\,\mathrm{Mpc}$ and $d = 5\,\mathrm{Mpc}$). The long–dashed line corresponds to $\gamma = 1.76$ and $r_o = 11\,\mathrm{Mpc}$.

**Figure 3.** The coefficients $\eta$, $\tau$ and $\zeta$ are plotted as functions of $\gamma$. For the cell–count analysis (Fig. 3a) they are evaluated inside cubic cells of unity size and for neighbor–count analysis (Fig.3b) inside spheres of unity radius. Error bars are standard deviations over ten different Montecarlo integrations.

**Figure 4.** The variance-skewness (triangles) and the variance-kurtosis (squares) relation for the CHDM runs (4a) and for CDM with $b = 1.5$ and $b = 1$ (4b) at the present time. On the left panels data on the DM particle distributions are plotted, while on the right we show results for galaxies, identified as high density peaks, with mean separation $d = 9$ Mpc. Errorbars are $1\sigma$ scatter over 20 bootstrap resamplings, while the dotted lines are the best fit of data to the hierarchical model $\bar{\xi}_q = S_q \bar{\xi}_2^{q-1}$.

**Figure 5.** The skewness and kurtosis coefficients ($S_3$ and $S_4$, respectively) for the distribution of DM particles, plotted in log units as functions of the scale. Triangles are for $S_3$ and squares are for $S_4$. Filled symbols are for analysis in real space, while the open ones refer to redshift space. Note that $S_3$ and $S_4$ must be independent of the scale as long as the hierarchical model holds. Data for the two runs of CHDM and for CDM at present time ($b = 1.5$) and at a more evolved stage ($b = 1$) are plotted.

**Figure 6.** The scale dependence of the CHDM $S_3$ coefficient in linear scales, for DM particles (circles) and for galaxies having $d = 5$ Mpc (squares) and $d = 9$ Mpc (triangles). Fig. 6a is for the CHDM runs, while Fig.6b is for CDM models. In Fig. 6a, upper panels are for the CHDM$_1$ run and the lower ones are for the CHDM$_2$ run. Left and right panels are for analyses in real and redshift space, respectively. In Fig. 6b upper panels are for CDM at present time ($b = 1.5$) and while lower panels are for the more evolved configuration at $b = 1$.

**Figure 7.** The scale–dependence of the $S_4$ coefficient in real space. The symbols are the same as in Figure 6.

**Figure 8.** Dynamical evolution of high–peak clustering for CHDM$_1$ and CHDM$_2$. Squares refer to the present time, triangles to $z = 1$.

**Figure 9.** $Q$ and $Q^{(4)}$, for $d = 5$ Mpc, from cells (circles) and neighbours (triangles), for real

and redshift space. Open circles refer to the analysis of PPS with neighbour technique. Error bars correspond to 3–$\sigma$'s. Let us remind that PPS data were tentatively corrected for redshift space distortions (see text).

**Figure 10.** Effects of peculiar velocities on clustering. The plots show galaxies with $d = 5$ Mpc in a square slice of 40 Mpc thickness cut out of the 100 Mpc simulation box. Fig. 10a is for the CHDM$_1$ model, while Figs. 9b and 9c are for the CDM model with $b = 1.5$ and $b = 1$, respectively. The slice is chosen to pass through a high density concentration and periodicity conditions are used to draw points outside the box boundaries. In redshift space (left pannels) the observer is placed on the $z = constant$ middle plane of the slice at position $x = 20$ and $y = 20$.

**Figure 11.** Direct comparison in real space among CHDM$_1$ (full circles), CHDM$_2$ (open circles) and PPS (triangles) based on neighbour counts. Fig. 11a is for the $Q$ coefficient, while Fig. 11b is for $Q^{(4)}$.

**Figure 12.** The deviation from the Poissonian VPF, $\sigma$, plotted against the scaling variable $N_c$, for the galaxies with $d = 9$ Mpc both in real (Fig. 12a) and in redshift (Fig. 12b) space. Different curves are for the hierarchical models described in §3.3; long–dashed for the hierarchical Poisson model, short–dashed for the negative binomial model, dotted for the phenomenological model of eq.(26) with $\omega$ parameters given by the best–fitting values reported in Table 4.

Table 1: Best–fit parameters for the 2–point correlation function $\xi(r) = (r_o/r)^\gamma$ evaluated in real space.

| | $r_o$ | $\gamma$ | $r_o$ | $\gamma$ |
|---|---|---|---|---|
| | CHDM$_1$ | | CHDM$_2$ | |
| DM particles | $10.4 \pm 0.5$ | $1.65 \pm 0.06$ | $8.1 \pm 0.2$ | $1.69 \pm 0.03$ |
| Galaxies d=5 | $14.8 \pm 0.7$ | $1.92 \pm 0.07$ | $10.1 \pm 0.5$ | $2.10 \pm 0.06$ |
| Galaxies d=9 | $17.3 \pm 1.7$ | $1.91 \pm 0.13$ | $10.8 \pm 0.8$ | $2.22 \pm 0.10$ |
| | CDM1.5 | | CDM1 | |
| DM particles | $8.1 \pm 0.6$ | $2.18 \pm 0.09$ | $22.4 \pm 1.6$ | $2.14 \pm 0.09$ |
| Galaxies d=5 | $7.2 \pm 0.5$ | $2.21 \pm 0.10$ | $9.7 \pm 0.3$ | $2.07 \pm 0.04$ |
| Galaxies d=9 | $8.0 \pm 1.0$ | $2.09 \pm 0.04$ | $10.3 \pm 0.8$ | $2.11 \pm 0.12$ |

Table 2: Best–fit values for reduced skewness ($S_3$) and kurtosis ($S_4$) for DM and galaxy distribution.

| | Real sp. | | Redshift sp. | |
|---|---|---|---|---|
| | $S_3$ | $S_4$ | $S_3$ | $S_4$ |
| **CHDM$_1$** | | | | |
| DM particles | $8.9 \pm 4.2$ | $166 \pm 150$ | $17.3 \pm 16.2$ | ...... |
| Galaxies d=5 | $3.2 \pm 0.5$ | $14.7 \pm 5.2$ | $3.0 \pm 0.7$ | $13.1 \pm 6.1$ |
| Galaxies d=9 | $2.6 \pm 0.3$ | $8.7 \pm 2.2$ | $2.5 \pm 0.3$ | $7.9 \pm 2.1$ |
| | | | | |
| **CHDM$_2$** | | | | |
| DM particles | $8.6 \pm 3.7$ | $167 \pm 163$ | $5.0 \pm 2.1$ | ...... |
| Galaxies d=5 | $2.8 \pm 0.5$ | $12.0 \pm 4.4$ | $2.9 \pm 0.4$ | $12.8 \pm 4.0$ |
| Galaxies d=9 | $2.3 \pm 0.2$ | $7.3 \pm 1.7$ | $2.5 \pm 0.3$ | $8.6 \pm 1.9$ |
| | | | | |
| **CDM1.5** | | | | |
| DM particles | $5.1 \pm 1.6$ | $47.2 \pm 30.5$ | $3.3 \pm 1.2$ | ...... |
| Galaxies d=5 | $3.0 \pm 0.3$ | $13.4 \pm 2.9$ | $3.2 \pm 0.4$ | $16.1 \pm 4.4$ |
| Galaxies d=9 | $2.7 \pm 0.3$ | $10.0 \pm 2.9$ | $2.4 \pm 0.1$ | $8.2 \pm 0.7$ |
| | | | | |
| **CDM1** | | | | |
| DM particles | $5.3 \pm 2.4$ | $57.0 \pm 61.0$ | $3.7 \pm 1.3$ | ...... |
| Galaxies d=5 | $2.8 \pm 0.2$ | $10.7 \pm 2.4$ | $3.2 \pm 0.3$ | $14.7 \pm 2.9$ |
| Galaxies d=9 | $2.3 \pm 0.2$ | $6.9 \pm 1.4$ | $2.4 \pm 0.1$ | $7.5 \pm 1.2$ |

Table 3: Values of the $S_3$ and $S_4$ coefficients in real space for our "bright" galaxies and for real galaxy samples. Only the Bouchet et al. (1993) result refer to redshift space.

| Sample | $S_3$ | $S_4$ |
|---|---|---|
| CHDM$_1$ | $2.6 \pm 0.3$ | $8.7 \pm 2.2$ |
| CHDM$_2$ | $2.3 \pm 0.2$ | $7.3 \pm 1.7$ |
| CDM1.5 | $2.7 \pm 0.3$ | $10.0 \pm 2.9$ |
| CDM1 | $2.3 \pm 0.2$ | $6.9 \pm 1.4$ |
| CfA (Peebles 1980) | $2.4 \pm 0.2$ | ...... |
| CfA (Fry & Gatzañaga 1993) | $2.0 \pm 0.2$ | $6.3 \pm 1.6$ |
| SSRS (Fry & Gatzañaga 1993) | $1.8 \pm 0.2$ | $5.4 \pm 0.2$ |
| IRAS (Meiksin et al. 1993) | $2.2 \pm 0.2$ | $10 \pm 3$ |
| IRAS (Fry & Gatzañaga 1993) | $2.2 \pm 0.3$ | $9.2 \pm 3.9$ |
| IRAS (Bouchet et al. 1993) | $1.5 \pm 0.5$ | $4.4 \pm 3.7$ |
| APM (Gatzañaga 1993) | $3.8 \pm 0.4$ | $33 \pm 7$ |

Table 4: The scaling parameter $\omega$ for the phenomenological expression (26) of the deviation from Poissonian VPF, $\sigma(N_c)$.

| Sample | $\omega$ Real space | $\omega$ Redsh. space |
|---|---|---|
| CHDM$_1$ | $0.41 \pm 0.05$ | $0.44 \pm 0.04$ |
| CHDM$_2$ | $0.46 \pm 0.03$ | $0.47 \pm 0.03$ |
| CDM1.5 | $0.47 \pm 0.05$ | $0.49 \pm 0.04$ |
| CDM1 | $0.48 \pm 0.04$ | $0.47 \pm 0.04$ |
| CfA (Alimi et al. 1990) | | $0.5 \pm 0.1$ |
| SSRS (Maurogordato et al. 1992) | | $0.6 \pm 0.1$ |
| IRAS (Bouchet et al. 1993) | | $0.9$ |

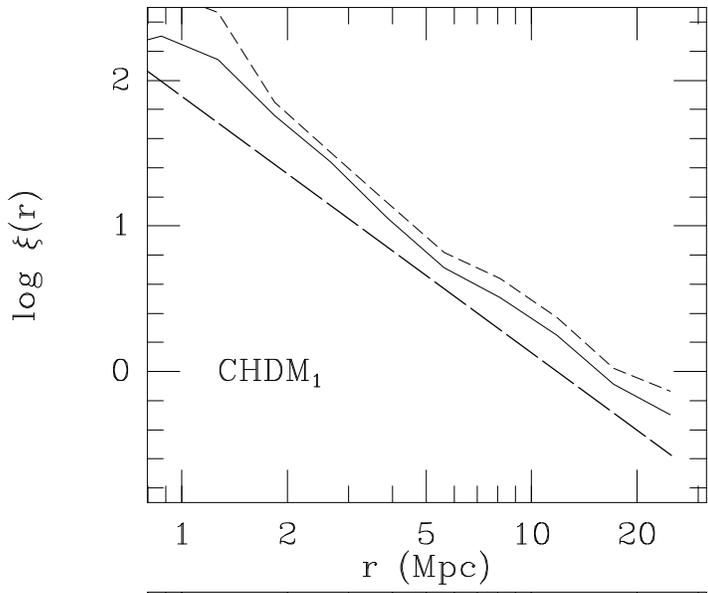
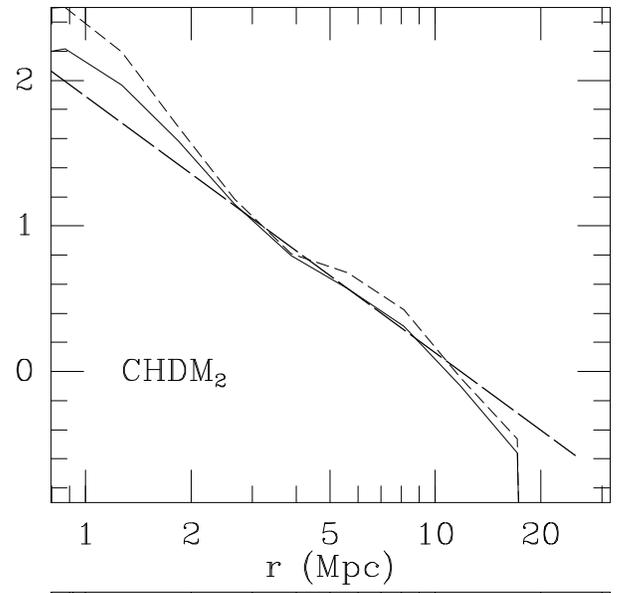
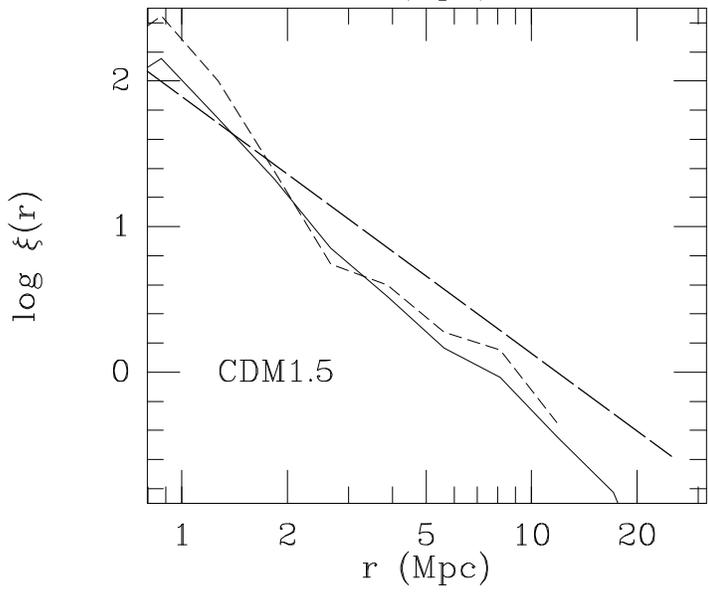
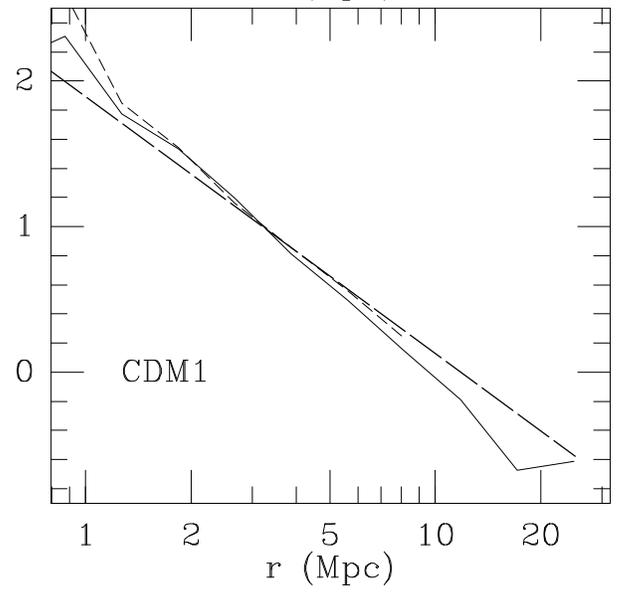

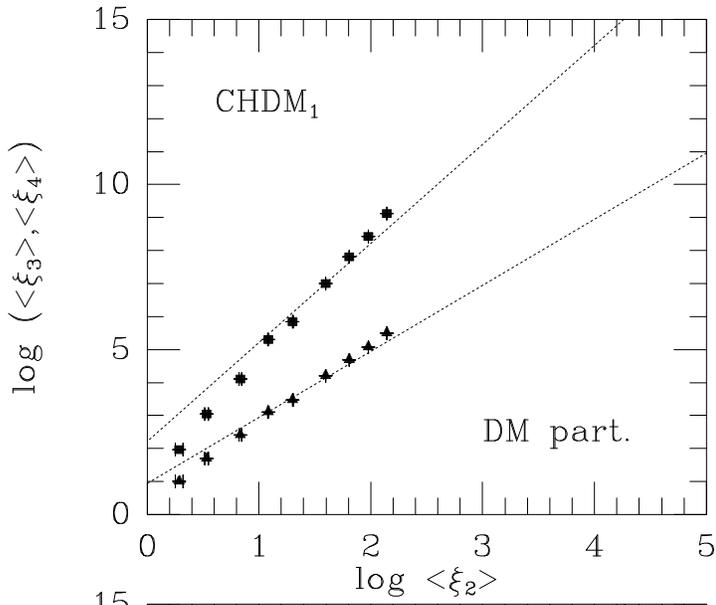
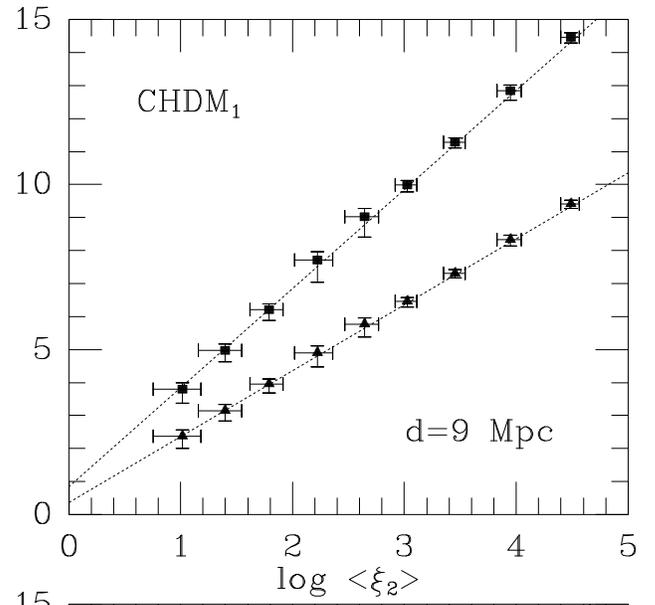
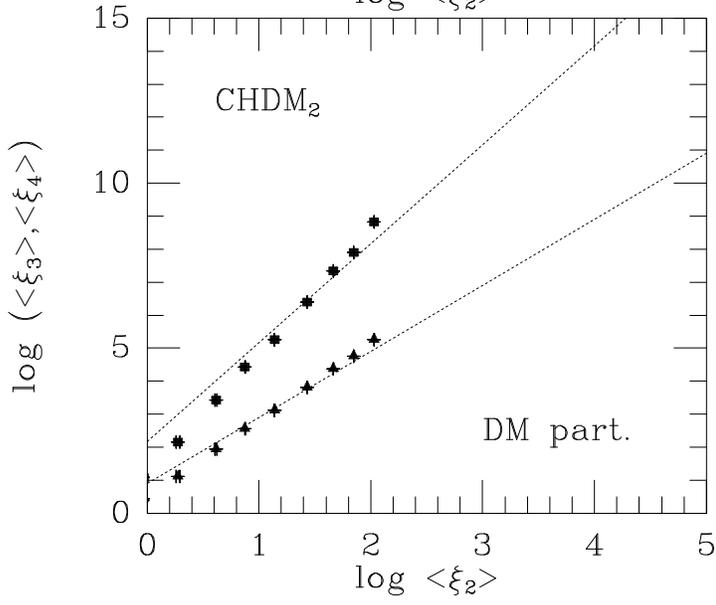
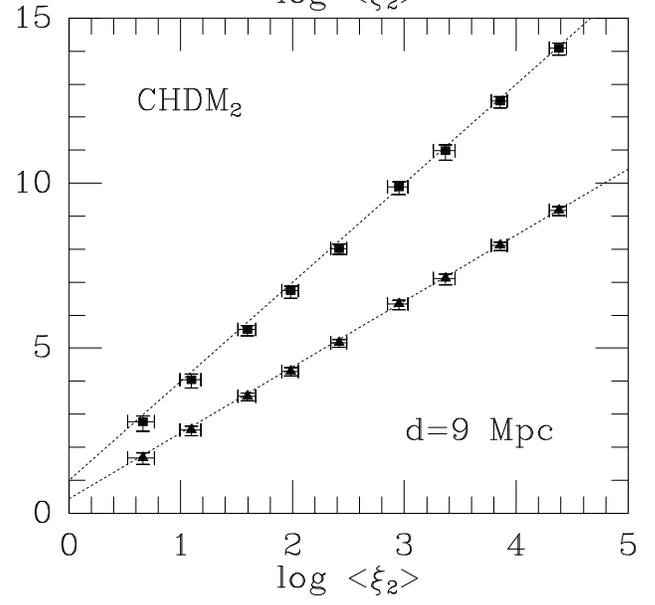

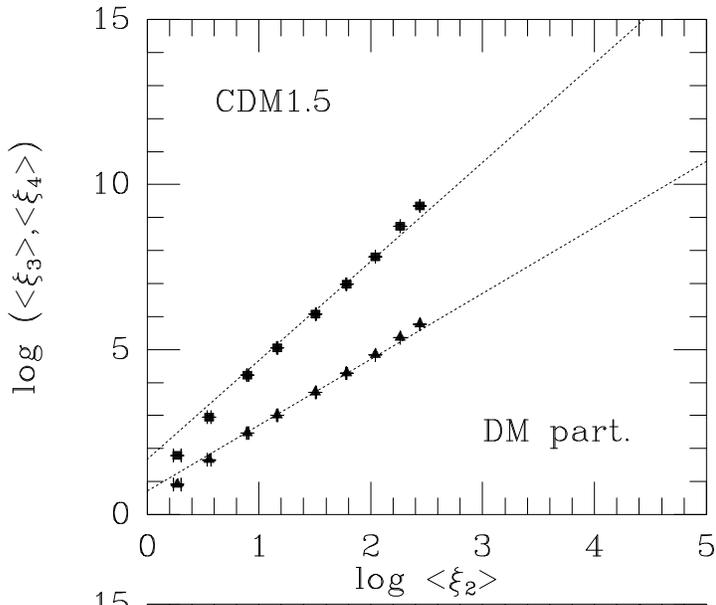
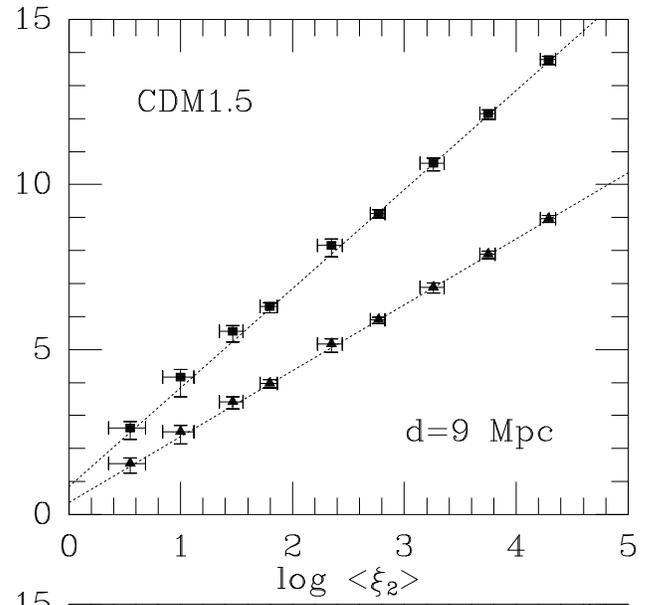
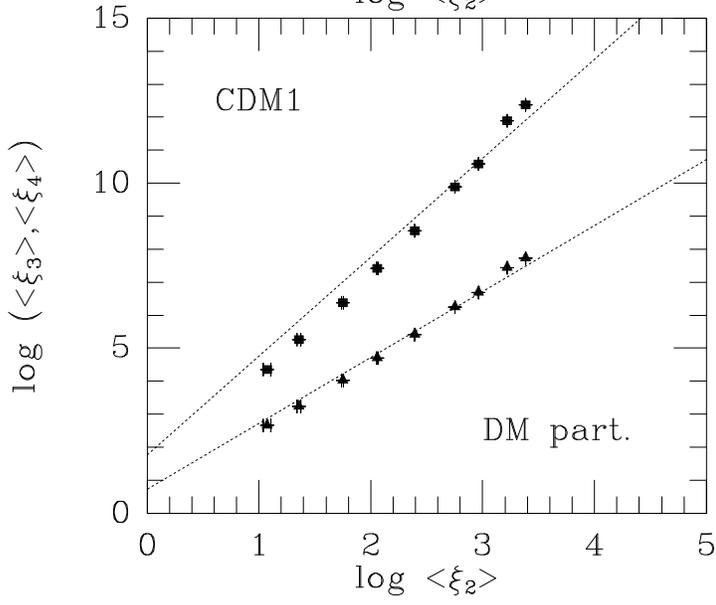
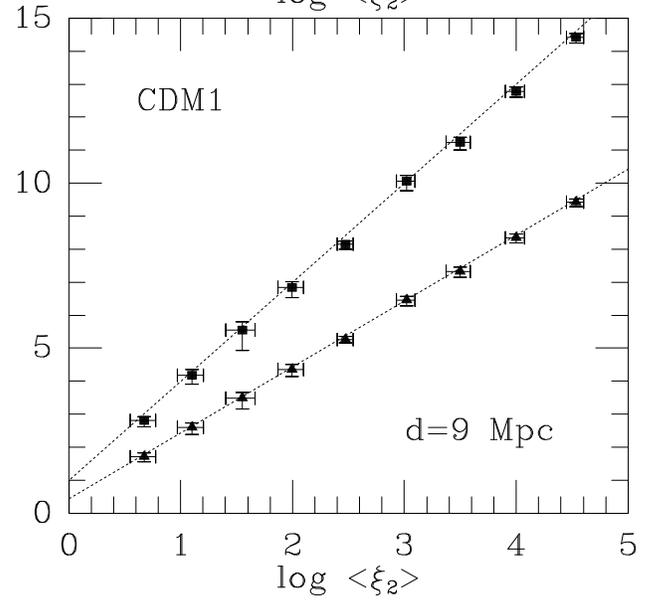

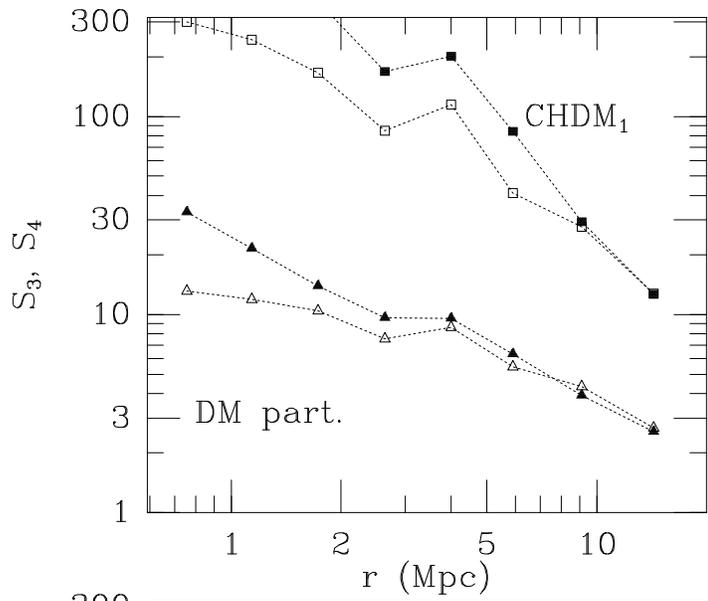
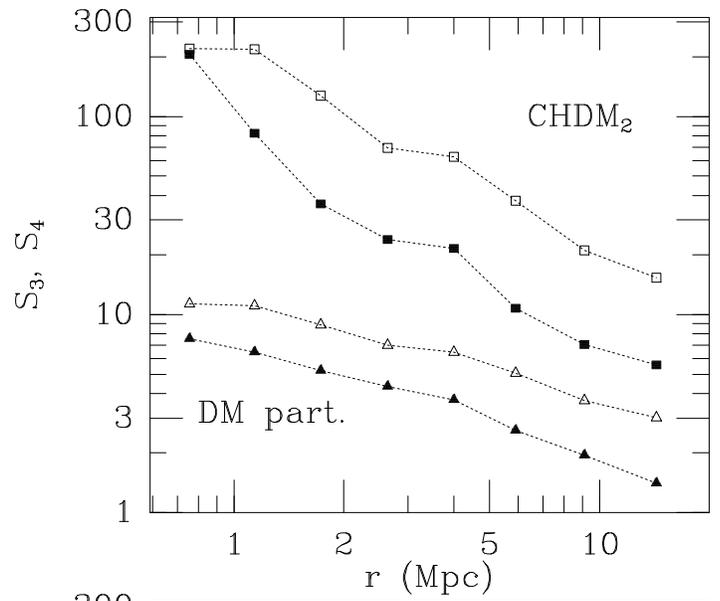
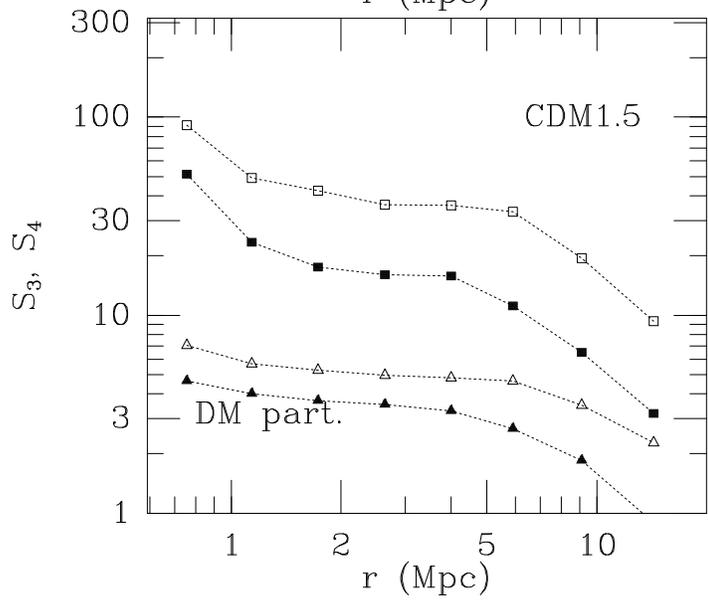
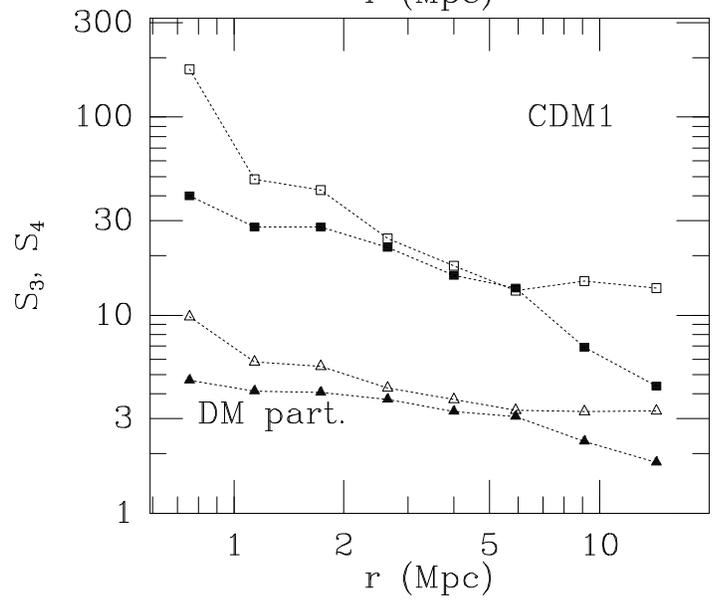

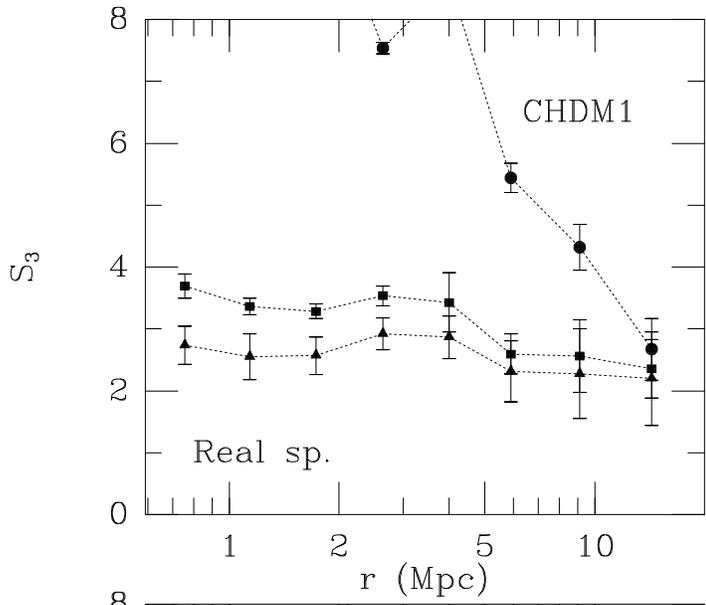
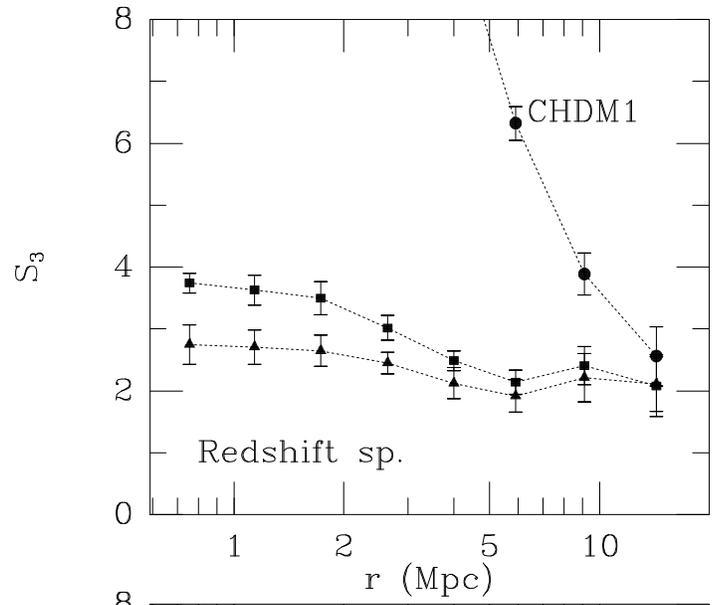
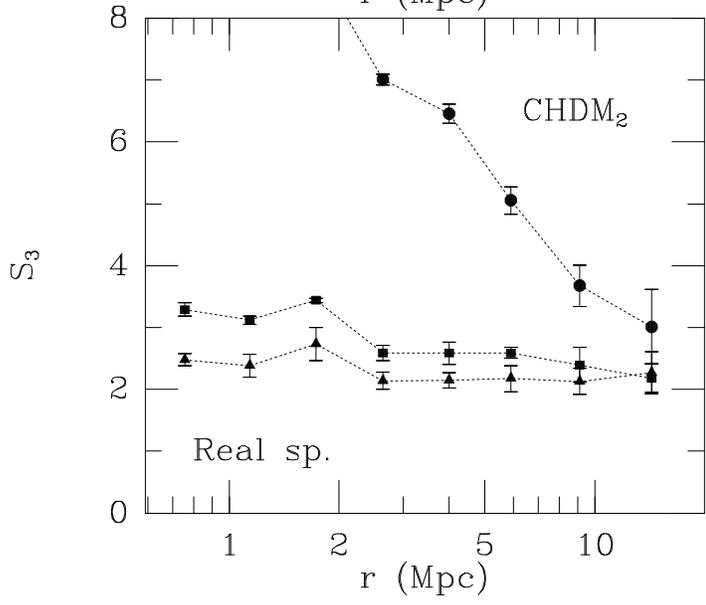
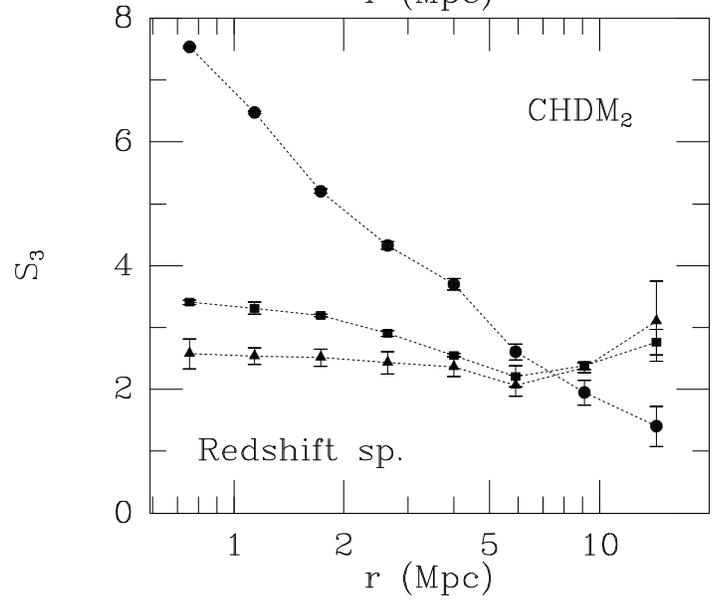

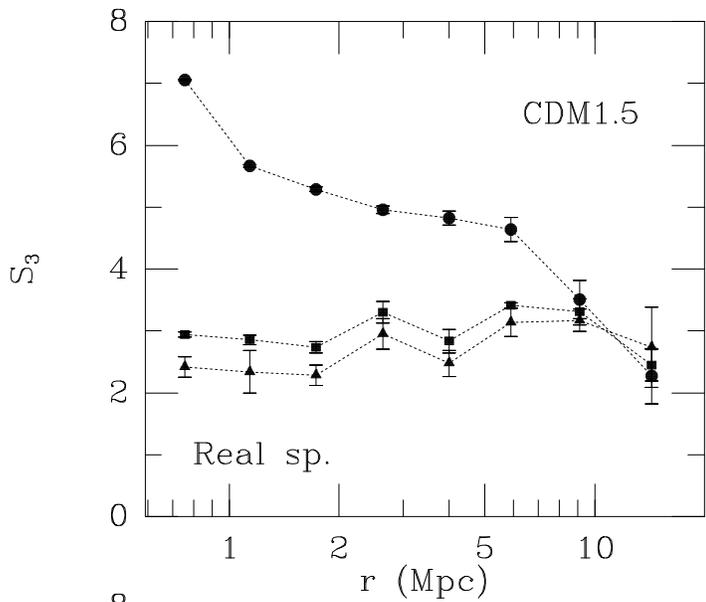
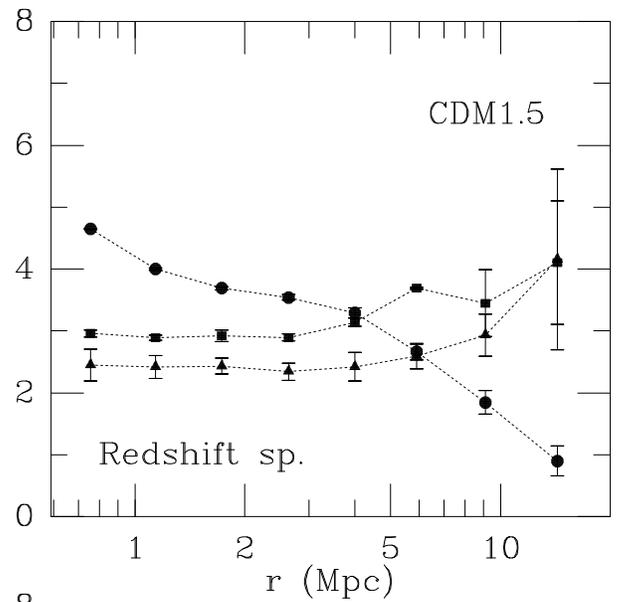
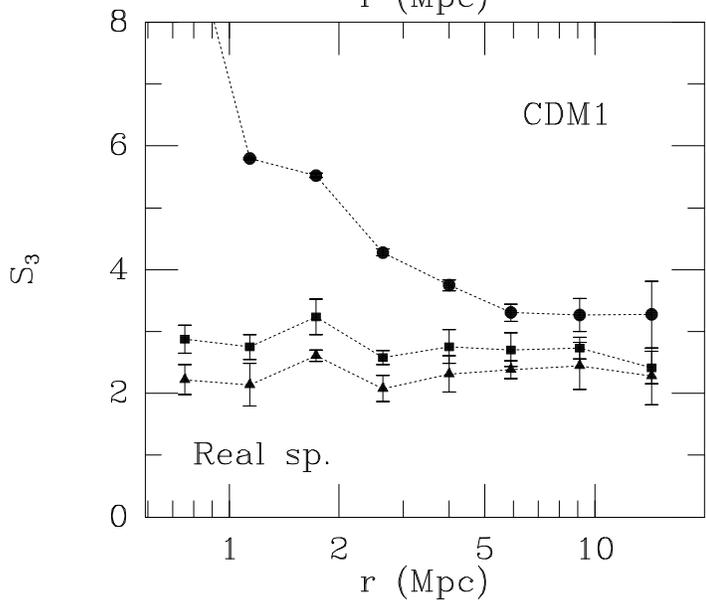
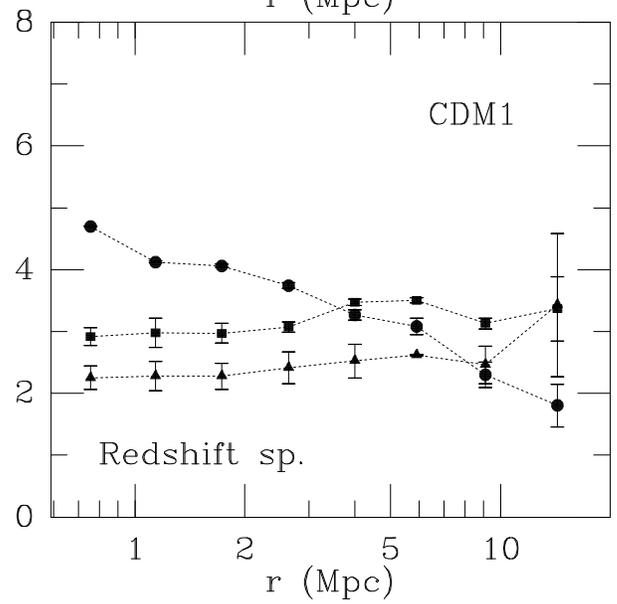

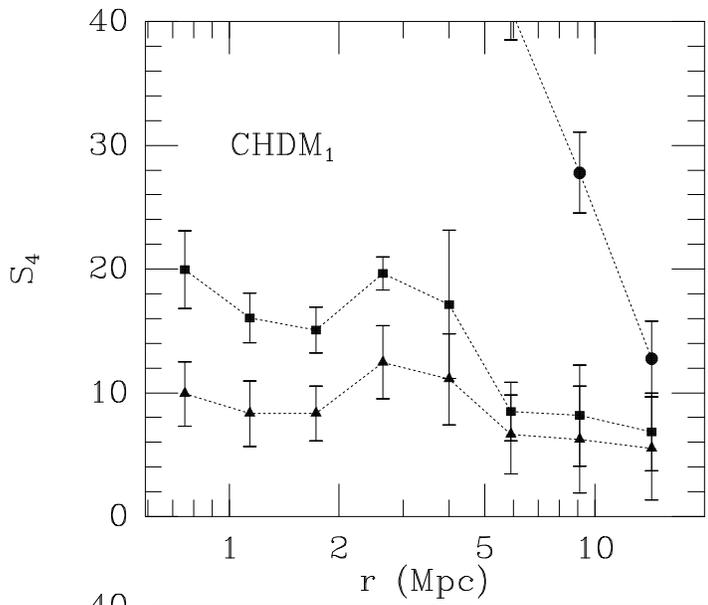
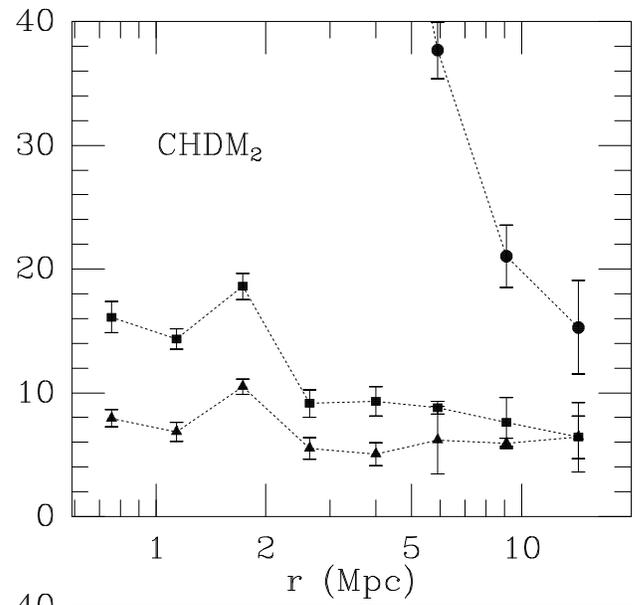
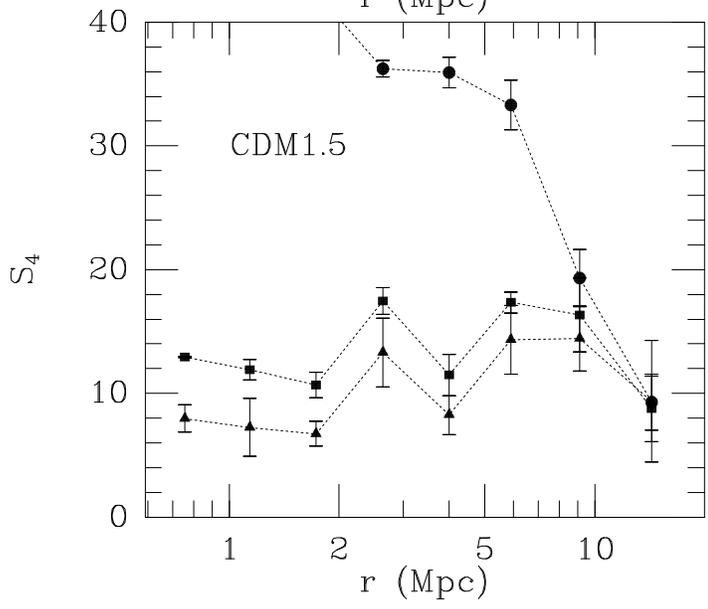
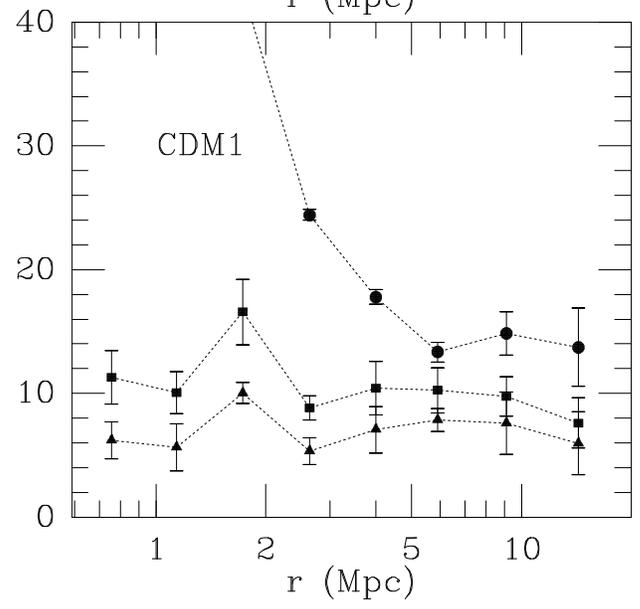

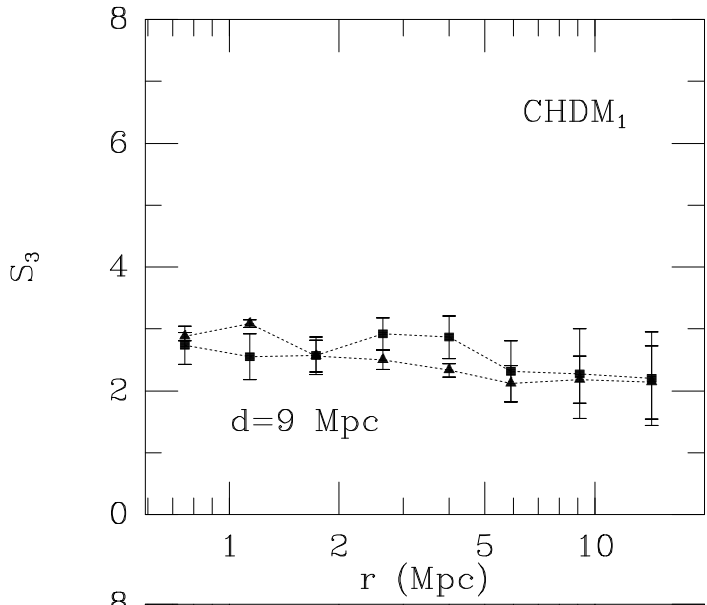
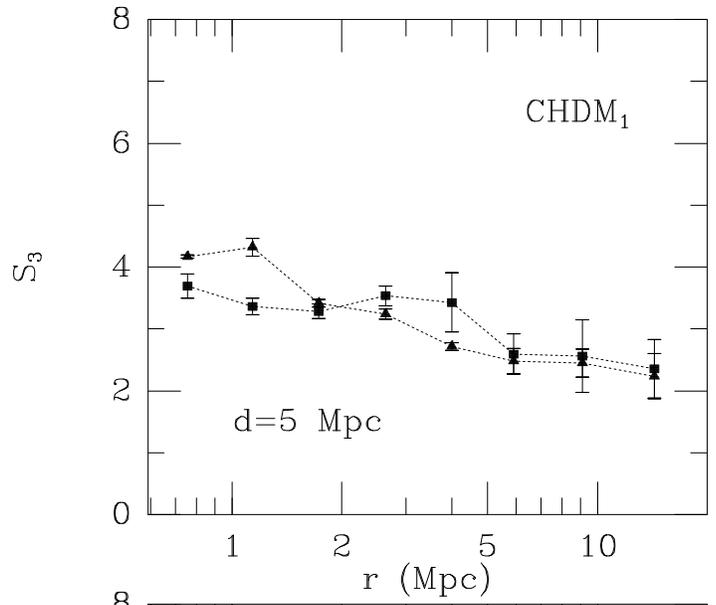
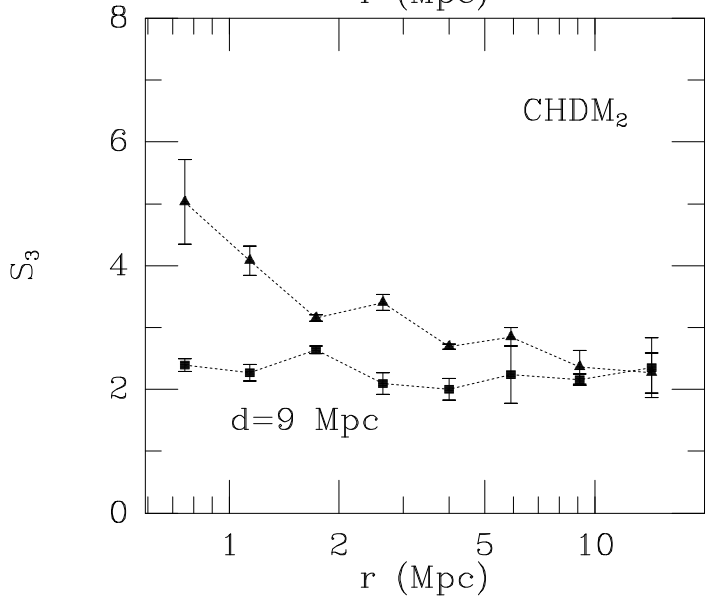
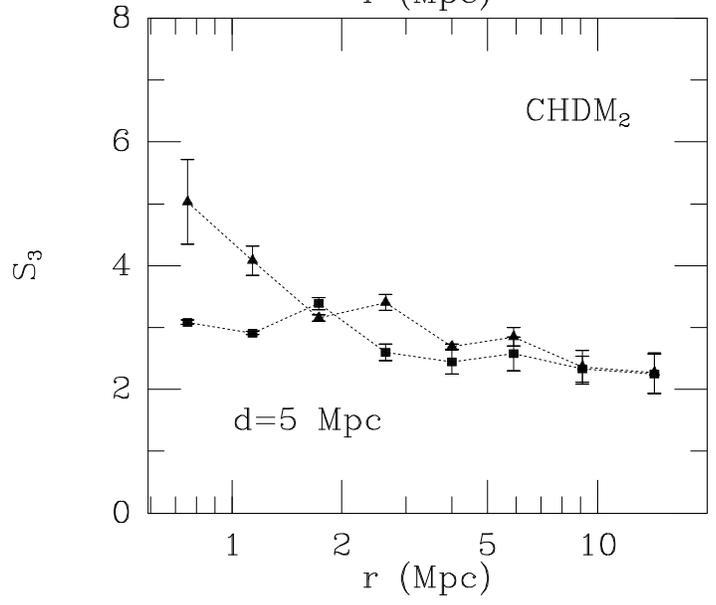

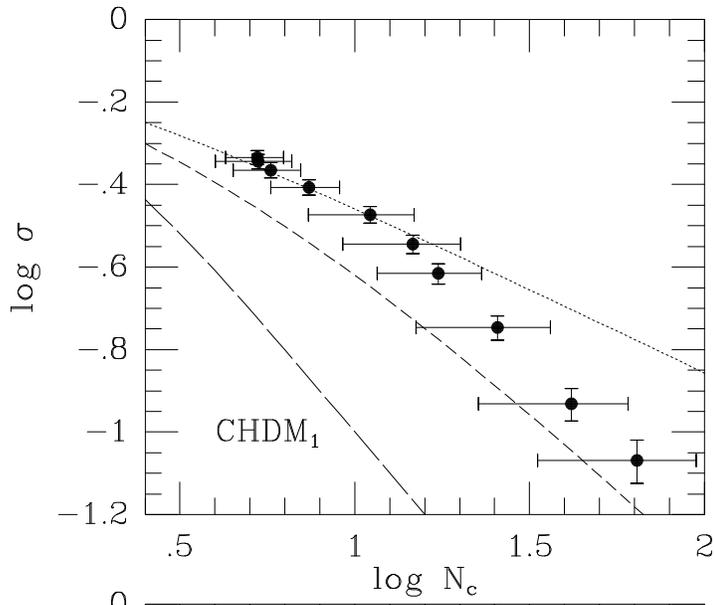
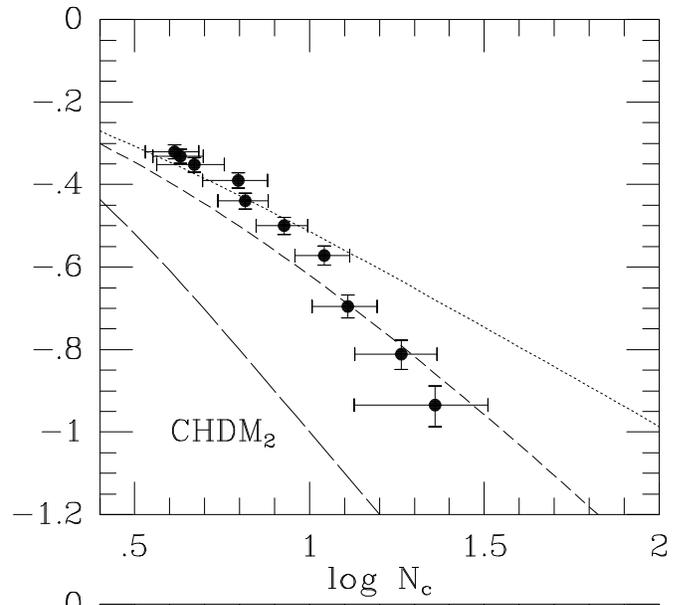
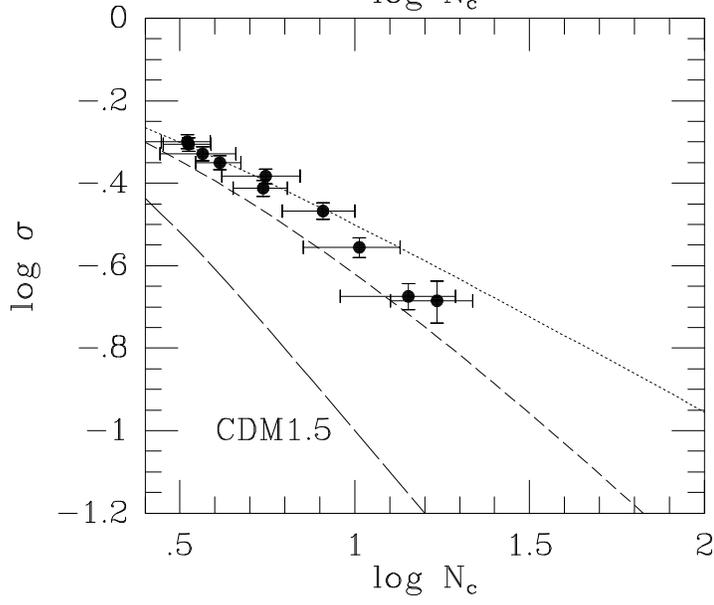
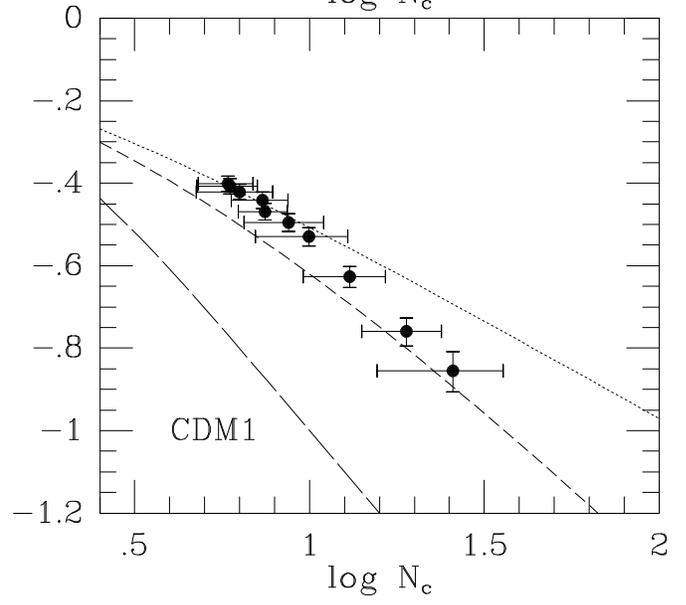

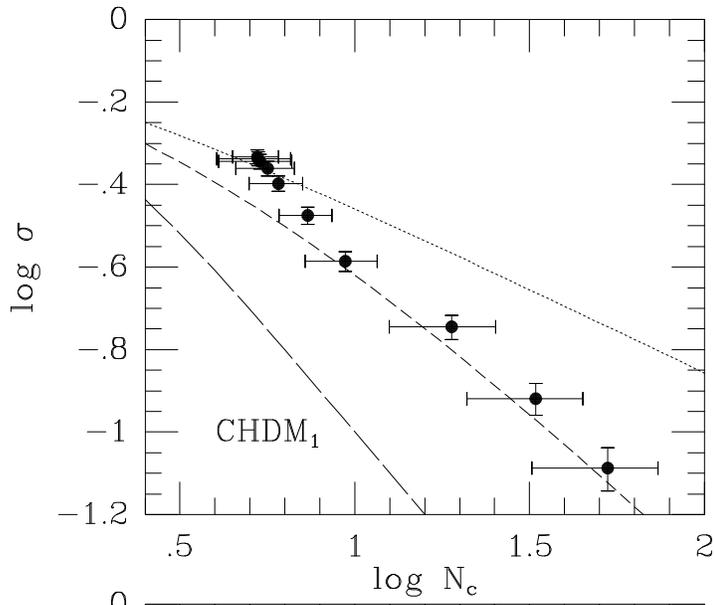
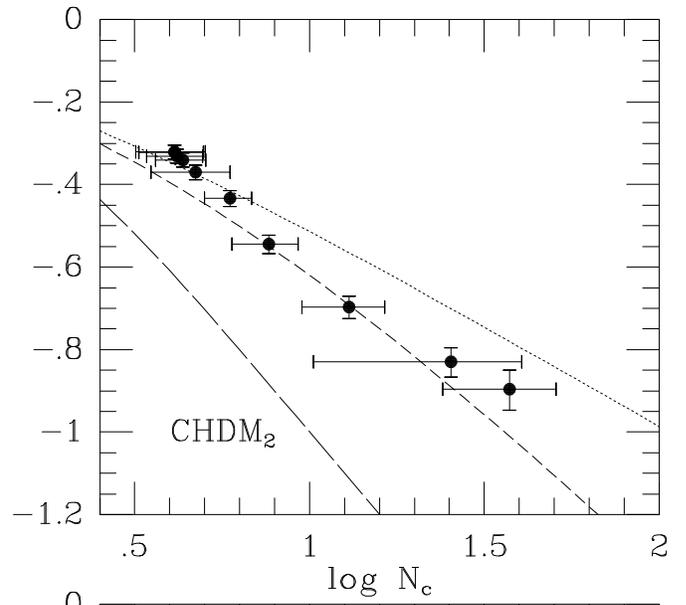
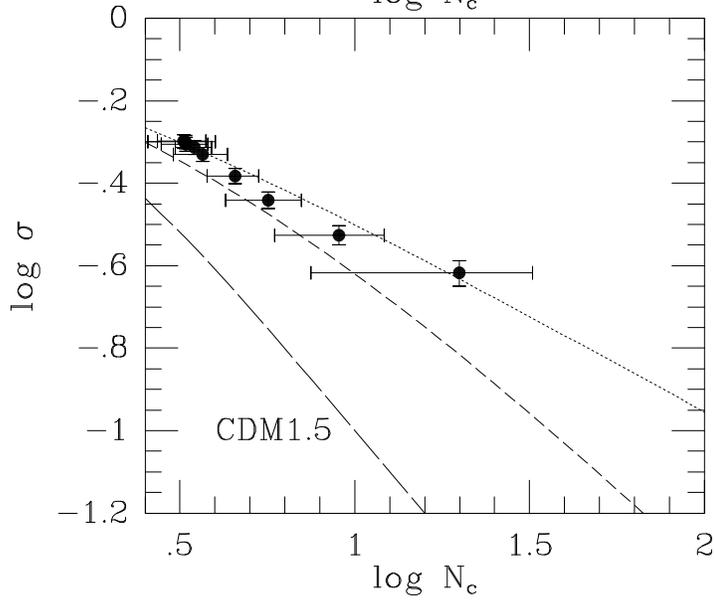
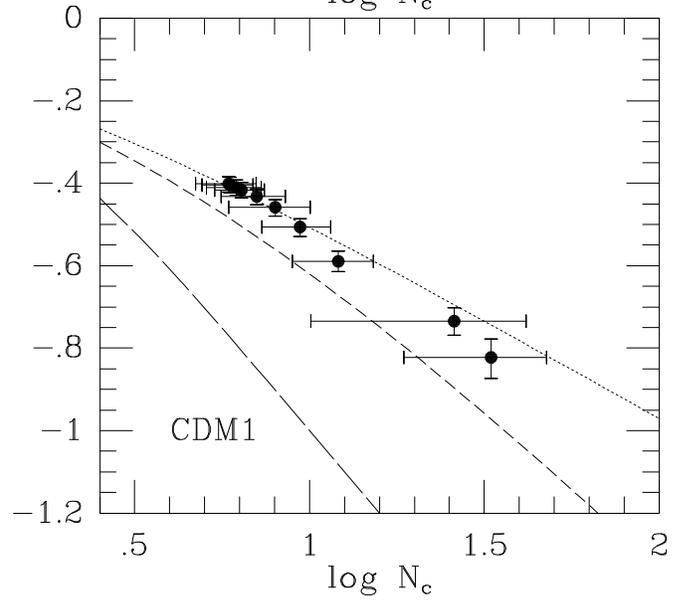